\documentclass{aa}  

\usepackage{graphicx}
\usepackage{txfonts}
\usepackage{pdflscape}

\begin{document}

\title{Dynamical traceback age of the Octans young stellar association\thanks{Tables A.1-A.4 are only available in electronic form at the CDS via anonymous ftp to cdsarc.u-strasbg.fr (130.79.128.5) or via http://cdsweb.u-strasbg.fr/cgi-bin/qcat?J/A+A/.}}

\subtitle{}

\author{
P. A. B. Galli\inst{1}
\and
N. Miret-Roig\inst{2}
\and
H. Bouy\inst{3}
\and
J. Olivares\inst{4}
\and
D. Barrado\inst{5}
}

\institute{N\'ucleo de Astrof\'isica Te\'orica, Universidade Cidade de S\~ao Paulo, R. Galv\~ao Bueno 868, 01506-000, S\~ao Paulo, SP, Brazil.\\ \email{phillip.ab.galli@gmail.com}
\and
University of Vienna, Department of Astrophysics, T\"urkenschanzstrasse 17, 1180 Wien, Austria.
\and
Laboratoire d'Astrophysique de Bordeaux, Univ. Bordeaux, CNRS, B18N, all\'ee Geoffroy Saint-Hilaire, 33615 Pessac, France.
\and
Depto. de Inteligencia Artificial, UNED, Juan del Rosal, 16, 28040 Madrid, Spain.
\and
Centro de Astrobiolog\'ia (CSIC-INTA), Camino Bajo del Castillo s/n, 28692 Villanueva de la Ca\~nada, Madrid, Spain.
}

\date{Received September 15, 1996; accepted March 16, 1997}

 \abstract
{Octans is one of the most distant ($d\sim150$~pc) young stellar associations of the solar neighbourhood, and it has not yet been sufficiently explored. Its age is still poorly constrained in the literature and requires further investigation.}
{We take advantage of the state-of-the-art astrometry delivered by the third data release of the \textit{Gaia} space mission combined with radial velocity measurements obtained from high-resolution spectroscopy to compute the 3D positions and 3D spatial velocities of the stars and derive the dynamical traceback age of the association. }
{We created a clean sample of cluster members by removing potential outliers from our initial list of candidate members. We then performed an extensive traceback analysis using different subsamples of stars, different metrics to define the size of the association, and different models for the Galactic potential to integrate the stellar orbits in the past. }
{We derive a dynamical age of $34^{+2}_{-2}$~Myr that is independent from stellar models and represents the most precise age estimate currently available for the Octans association. After correcting the radial velocity of the stars for the effect of gravitational redshift, we obtain a dynamical age of $33^{+3}_{-1}$~Myr, which is in very good agreement with our first solution. This shows that the effect of gravitational redshift is small for such a distant young stellar association. Our result is also consistent with the less accurate age estimates obtained in previous studies from lithium depletion (30-40~Myr) and isochrones (20-30~Myr). By integrating the stellar orbits in time, we show that the members of Octans and Octans-Near had different locations in the past, which indicates that the two associations are unrelated despite the close proximity in the sky. }
{This is the first reliable and precise dynamical age result for the Octans young stellar association. Our results confirm that it is possible to derive precise dynamical ages via the traceback method for $\sim30$~Myr old stellar clusters at about $\sim150$~pc with the same precision level that has been achieved in other studies for young stellar groups within 50~pc of the Sun. This represents one more step towards constructing a self-consistent age scale based on the 3D space motion of the stars in the young stellar clusters of the solar neighbourhood. }
\maketitle

\section{Introduction}

Stellar associations are sparsely populated groups of coeval and co-moving stars that contain a few dozen to hundreds of members \citep{Torres2008}. It is implicitly assumed that the members of a stellar association were born from the same star formation episode and therefore have similar properties. The close proximity and young age of the stellar associations in the solar neighbourhood has turned them into primary laboratories for investigating the properties of stars at early stages of stellar evolution \citep[see e.g.][]{2006AJ....131.3109G,2009A&A...495..523B} and to search for planetary-mass companions around young stars \citep[see e.g.][]{2004A&A...425L..29C,2010Sci...329...57L}. While many of the nearby young stellar associations have been intensively investigated in the past, some are still poorly characterized and deserve more study. 

The Octans association, located at about 150~pc, is one such young stellar association that has not yet been sufficiently explored in the literature, presumably due to its sky location far in the southern hemisphere (the declination of the stars ranges from -86 to -20~deg). Octans was first recognized as a young stellar association by \citet{Torres2003a,Torres2003b} after the discovery of six members (all of them young G type stars) around the south celestial pole. The subsequent review of \citet{Torres2008} increased the number of members to 15~stars.  Very few studies have targeted this young stellar association since. Some recent surveys failed to identify this association or incorrectly associated some of its members with other stellar groups \citep{2017AJ....153...95R,2022ApJ...939...94M}. One reason for this is the large spatial extent and sparsity of its members, which makes differentiating between cluster members and field stars more difficult.

One important step towards improving the census of Octans was taken by \citet{Murphy2015}, who performed the first survey of low-mass cluster members and identified 29~stars with spectral types ranging from K5-M4 based on spectroscopic observations. The census of the stellar population in Octans was further expanded with the BANYAN~$\Sigma$ tool, which implements a Bayesian classifier based on stellar positions and spatial velocities to compute membership probabilities of stars in the nearby young stellar associations of the solar neighbourhood \citep{2018ApJ...856...23G,2018ApJ...860...43G,2018ApJ...862..138G}. 

In contrast to the other young stellar associations of the solar neighbourhood, the age of Octans is also still poorly constrained in the literature. \citet{Torres2008} used the age of 10~Myr in their membership analysis to select additional cluster members and later assigned an a posteriori age estimate of 20~Myr based on the observed dispersion of the stellar positions. \citet{Murphy2015} identified the Li~I~$\lambda$6708 absorption line in nine stars of their sample, but unfortunately their spectroscopic survey did not span the location of the lithium depletion boundary (LDB), which prevented them from accurately deriving the age of the association using this method. By comparing the lithium depletion pattern of K and M type stars in Octans with other well-studied young stellar associations, the authors roughly estimate an age range of 30 to 40~Myr. 

The third data release of the \textit{Gaia} space mission \citep[\textit{Gaia}-DR3;][]{2023A&A...674A...1G} provides 5D astrometry (positions, proper motions, and parallaxes) and radial velocity measurements for most stars in the Octans association. Now the dynamical age of the group can be inferred from the traceback method using precise information of the 3D positions and 3D spatial velocity of the stars derived from \textit{Gaia}-DR3 data. The traceback strategy is used to trace the stellar orbits of individual cluster members into the past and determine the time when the stars were in the closest proximity, which is assumed to be the birth time of the association \citep{2003ApJ...599..342S,2004ApJ...609..243O,2006AJ....131.2609D,2014A&A...563A.121D,MiretRoig2018}. The dynamical age is considered to be a semi-fundamental age because it involves making a few assumptions (e.g. choice of the Galactic potential to integrate the stellar orbits), but these assumptions do not affect the derived age to a large degree \citep{2010ARA&A..48..581S}. Our team has recently derived the dynamical age of the $\beta$~Pictoris moving group \citep{2020A&A...642A.179M}, various stellar groups of the Upper Scorpius and Ophiuchus star-forming regions\citep{2022A&A...667A.163M},  and the Tucana-Horologium association \citep{2023MNRAS.520.6245G} based on a refurbished version of the traceback method. A more recent study conducted by \citet{2023ApJ...946....6C} has addressed the various sources of systematic errors that exist in the dynamical ages derived via the traceback method. A variant of the traceback method based on a forward modelling of the stellar orbits has also been used to identify co-moving stars  and derive precise kinematic ages of young stellar associations \citep{2019MNRAS.489.3625C}. All these studies have demonstrated the feasibility of deriving precise age estimates that are independent of stellar models and consistent with the results from other age-dating methods.

This paper is one in a series dedicated to constructing a consistent age scale for the young stellar clusters of the solar neighbourhood. Here, we focus on the dynamical age and evolution in time of the Octans young stellar association. In Section~\ref{sect2} we introduce our sample of stars, present the data used in our analysis (including archival material and high-resolution spectra we collected ourselves), and describe our selection criteria for defining a clean sample of cluster members. In Section~\ref{section3} we discuss our strategy for inferring the dynamical age of the association based on the traceback method. In Section~\ref{section4} we discuss our results and compare them with other nearby young stellar clusters. Finally, in Section~\ref{section5} we summarize our conclusions.

\section{Sample and data}\label{sect2}

\subsection{Initial list of cluster members}

Our initial sample of Octans stars is based on the lists of cluster members identified from the BANYAN project \citep{2018ApJ...856...23G,2018ApJ...860...43G,2018ApJ...862..138G}, which provides the most complete censuses for most of the young local associations based on \textit{Gaia} data. The members compiled by \citet{Murphy2015} were taken into account to produce the BANYAN sample for the Octans association. This compilation of cluster members includes 103 stars after removing the sources in common among the different versions of the BANYAN catalogue. We counted the components of binaries and multiple systems that have been resolved in previous studies as independent entries in our sample. As explained below, we refined this initial sample of candidate members to select the most likely kinematic members of the association that will be used in the traceback analysis. 

\subsection{Proper motions and parallaxes}
We cross-matched our initial sample of Octans stars with the \textit{Gaia}-DR3 catalogue and we found proper motions and parallaxes for all sources in this sample. The median precision in proper motions and parallaxes is 0.016~mas/yr, 0.018~mas/yr, and 0.017~mas, respectively. This translates into a precision of about 0.023~km/s and 0.026~km/s in the 2D tangential velocity of the stars confirming that the tangential velocity of the stars can be derived with good precision thanks to the state-of-the-art precision of the \textit{Gaia} astrometry. The main contribution to the error budget of the 3D velocity of the stars comes from the radial velocity measurements that are currently available (see below).

\subsection{Radial velocities}\label{sect2_RV}

The \textit{Gaia}-DR3 catalogue provides radial velocities for 86~stars in the sample with a median precision of 0.7~km/s. Despite the relatively large number of cluster members with radial velocities available in our sample, we note that many of these sources exhibit a radial velocity precision of a few km/s. We proceeded as follows to increase the number of sources with radial velocity information and improve the precision of radial velocity measurements.

First, we searched for high-resolution spectra ($R\gtrsim 40\,000$) of the stars in our sample in public archives of several ground-based facilities. We found 76 archival spectra for the stars in our sample collected with different instruments as summarized in Table~\ref{tab1}. Second, we performed our own observations with the CHIRON optical high-resolution spectrograph \citep{2013PASP..125.1336T} mounted on the SMARTS 1.5m telescope at the Cerro Tololo Inter-American Observatory (CTIO). We observed 25 stars in service mode from April to November 2021 (programme ID: 661, PI: Bouy). 

\begin{table*}[!h]
\centering
\caption{Properties of the spectra downloaded from public archives.
\label{tab1}}
\begin{tabular}{ccccc}
\hline\hline
Instrument & R & $\Delta\lambda$ & Number of spectra& Programme identifier\\
&&(nm)&\\
\hline\hline
ESO/HARPS&115\,000&378-691&39&0103.C-0759, 106.21TJ\\
ESO/UVES&110\,000&300-1100&37&079.C-0556, 082.C-0218, 089.C-0207, \\
&&&&095.C-0437, 098.C-0463, 106.21S8, 110.23QM \\
\hline\hline

\end{tabular}
\tablefoot{We provide for each instrument the maximum resolving power, spectral range, number of spectra retrieved from the archives and programme identifier of the corresponding data.}
\end{table*}

We employed the same methodology to analyse the spectra we collected and those downloaded from the archives. Each spectrum is reduced with the pipeline of the corresponding instrument and the radial velocity is computed with iSpec \citep{2014A&A...569A.111B} as explained below. We cross-correlated the target spectrum with template spectra of different spectral types (A0, F0, G2, K0, K5, and M5) and computed the cross-correlation function in each case. The radial velocity of the target spectrum is given by the result obtained from the template with the closest spectral type to the target. The scatter in radial velocity computed from the three closest templates to the target  (i.e. the template with the closest spectral type to the target, one before and one after) is added in quadrature to the formal radial velocity uncertainty given by iSpec to account for the observed fluctuation in the results obtained from different templates. We inspected the cross-correlation functions individually and discarded the radial velocity solutions that result from a poor fit to the cross-correlation function (e.g. due to a low signal to noise ratio of the spectrum or a mismatch between the spectral type of the target and template). The radial velocities derived from each spectrum are given in Table~\ref{tabA1}.  

Many of the spectra downloaded from the archives are from the same target. We proceeded as follows to combine the various radial velocities of the (single) stars with multiple spectra in our sample. First, we used the solution derived from a single spectrum to generate 1\,000 synthetic measurements from a Gaussian distribution using the radial velocity value and its uncertainty as the mean and variance of the distribution, respectively.  Then, we used the joint distribution of the ensemble of synthetic radial velocity measurements produced by this method from all spectra available from the same target to derive the final radial velocity of the target. We took the mean of the joint distribution of radial velocity measurements and the 16\% and 84\% percentiles of the distribution to estimate the radial velocity of the star and its uncertainty. We provide the radial velocity of the stars we derived based on this strategy in Table~\ref{tabA2}. 

\begin{figure}
\begin{center}
\includegraphics[width=0.49\textwidth]{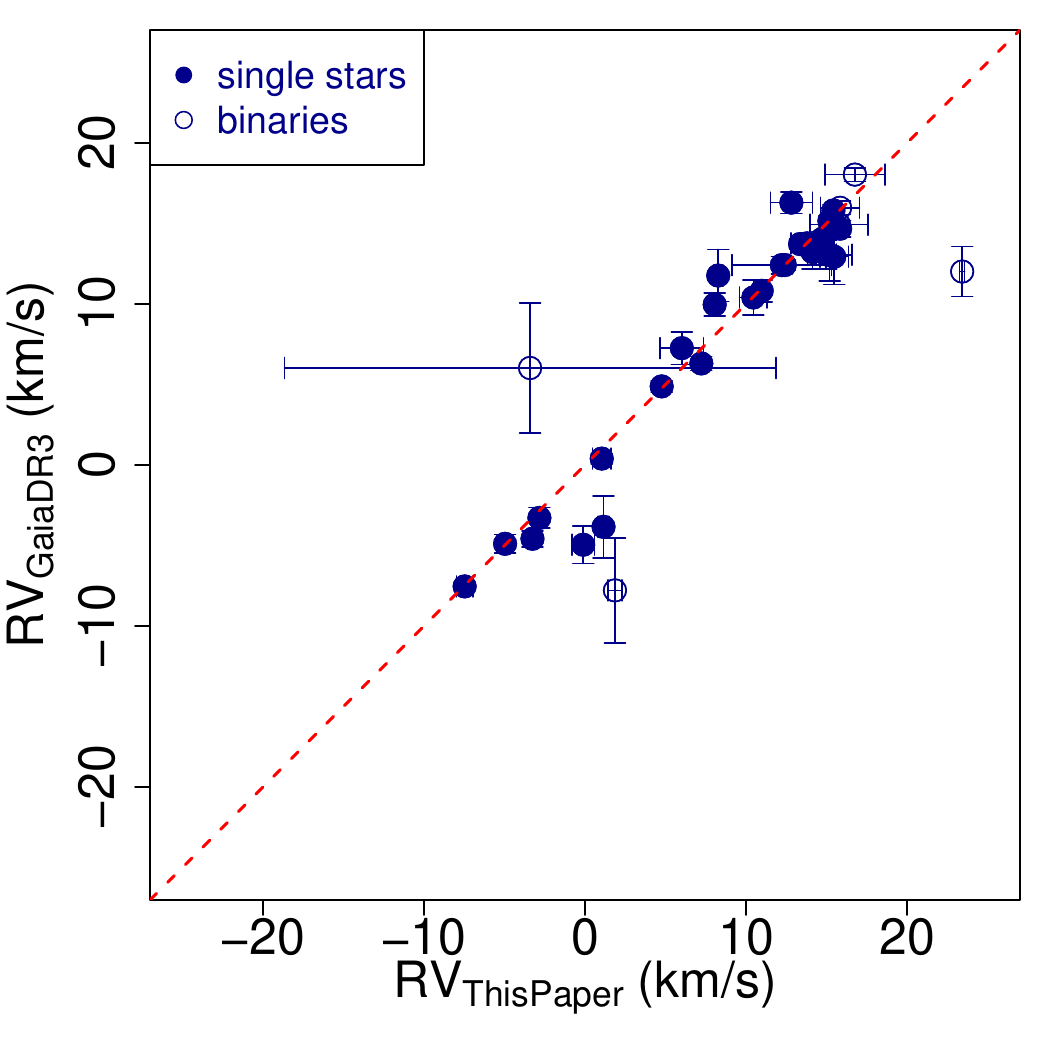}
\caption{
\label{fig_RV_comp} 
Comparison of the radial velocities derived in this paper with the \textit{Gaia}-DR3 catalogue. Open symbols denote known binaries identified in the literature.The dashed red line indicates a perfect correlation between the radial velocity measurements in the two datasets. 
}
\end{center}
\end{figure}

By combining our own observations with the archival spectra downloaded from public repositories we derived the radial velocity for 34~stars in this paper. Figure~\ref{fig_RV_comp} compares the radial velocities derived in this paper with the ones from the \textit{Gaia}-DR3 catalogue for the sources in common. We note that the radial velocities we derived are mostly more precise. The median radial velocity precision of the stars in our sample is 0.4~km/s while the median radial velocity precision for the same sources in the \textit{Gaia}-DR3 catalogue is 0.6~km/s. The mean difference in radial velocity (in the sense, this paper `minus' \textit{Gaia}-DR3) for the single stars in our sample is $0.4\pm0.4$~km/s and the root mean square error between the two datasets is 1.8~km/s.

\subsection{Youth diagnostics}\label{sect2_youth_diagnostics}

We searched the literature for youth indicators to investigate whether the stars in our sample show signs of youth. This analysis combined with the selection of the most likely kinematic members of the association (see Sect.~\ref{sect2_sample_selection}) defines a clean sample of Octans members that can be used for the traceback analysis. 

\begin{figure*}
\begin{center}
\includegraphics[width=0.49\textwidth]{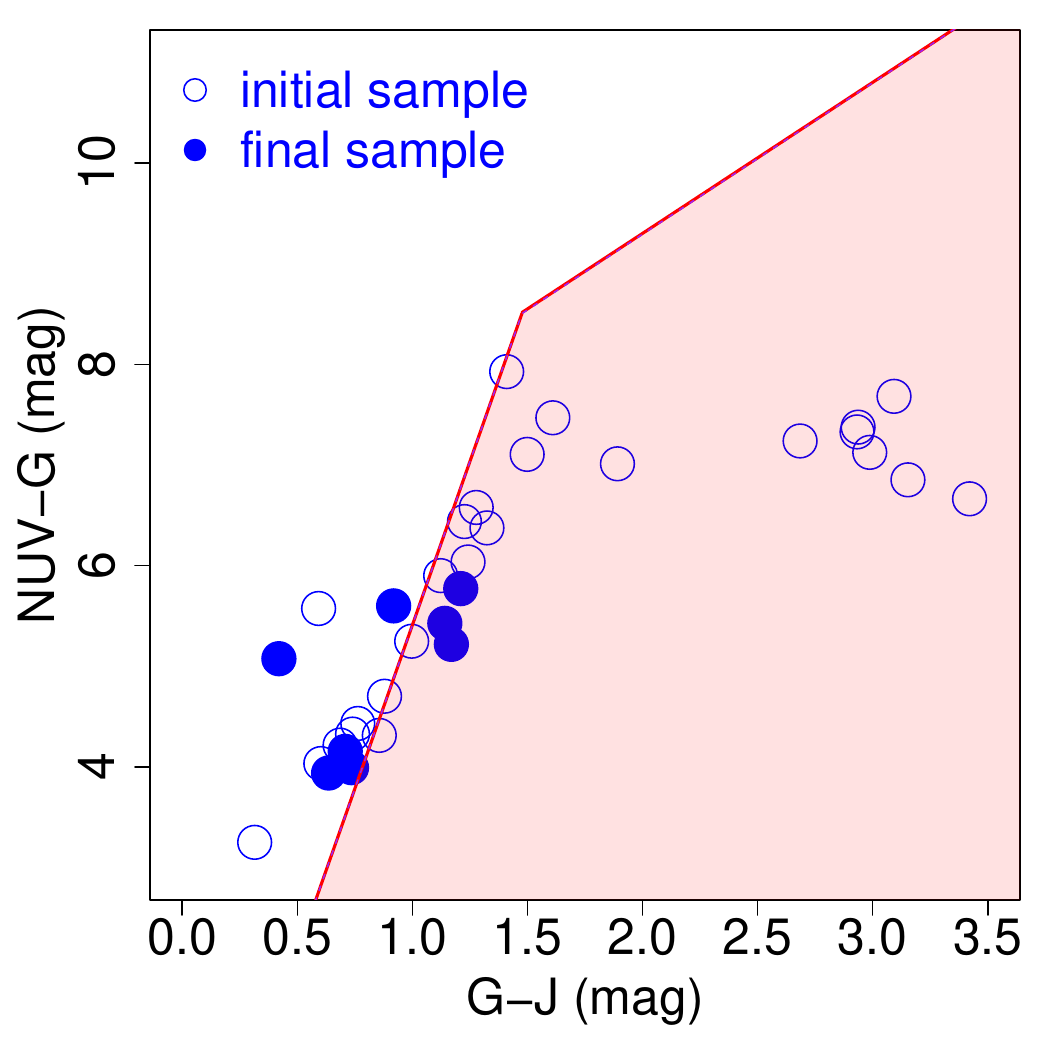}
\includegraphics[width=0.49\textwidth]{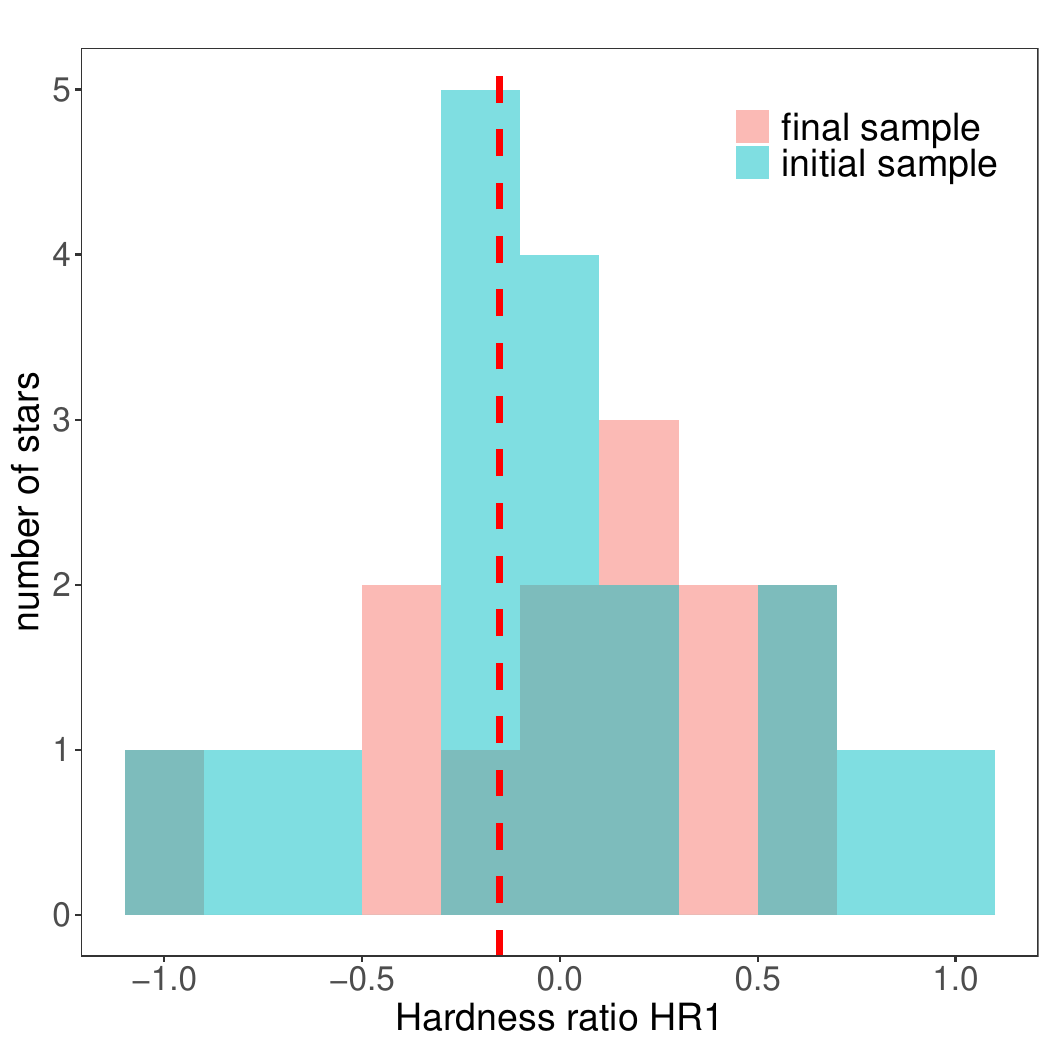}
\includegraphics[width=0.49\textwidth]{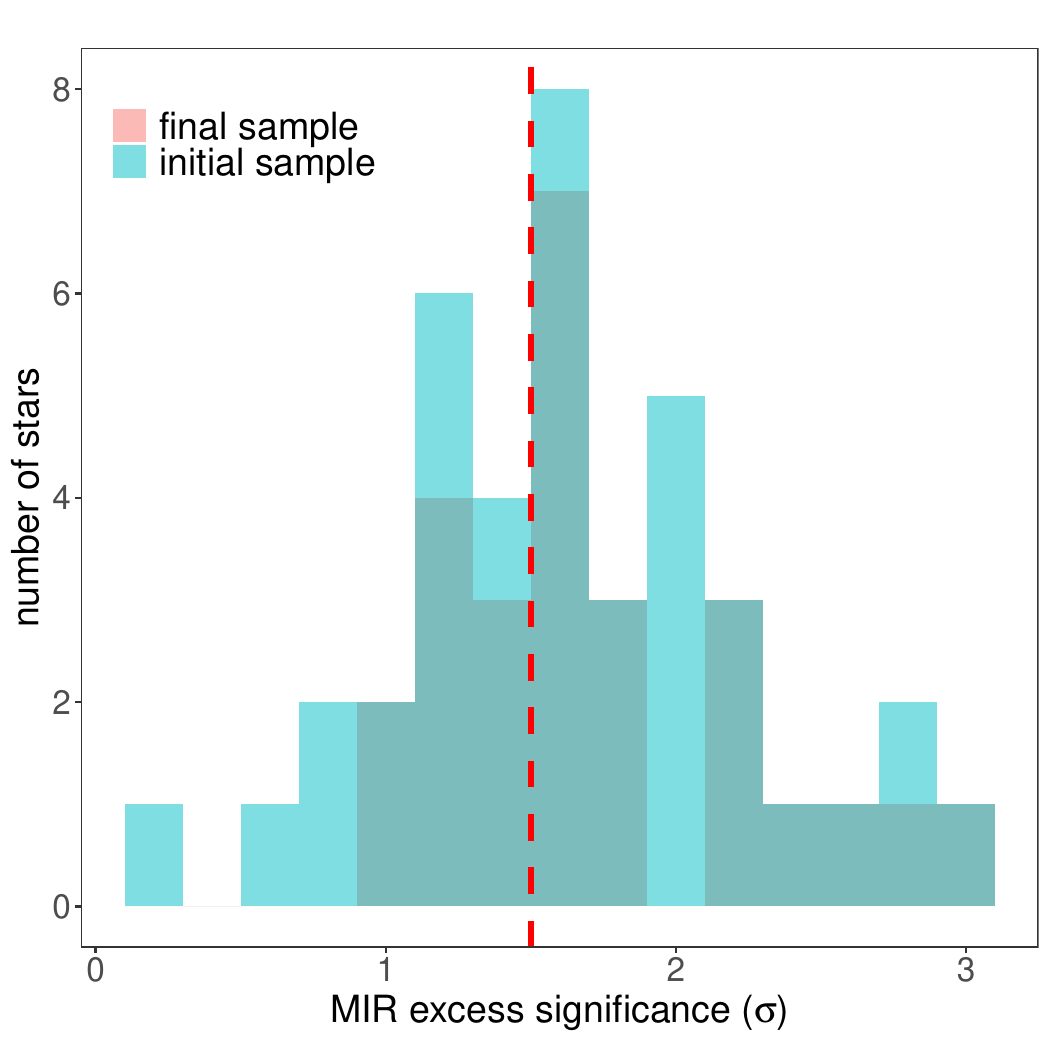}
\includegraphics[width=0.49\textwidth]{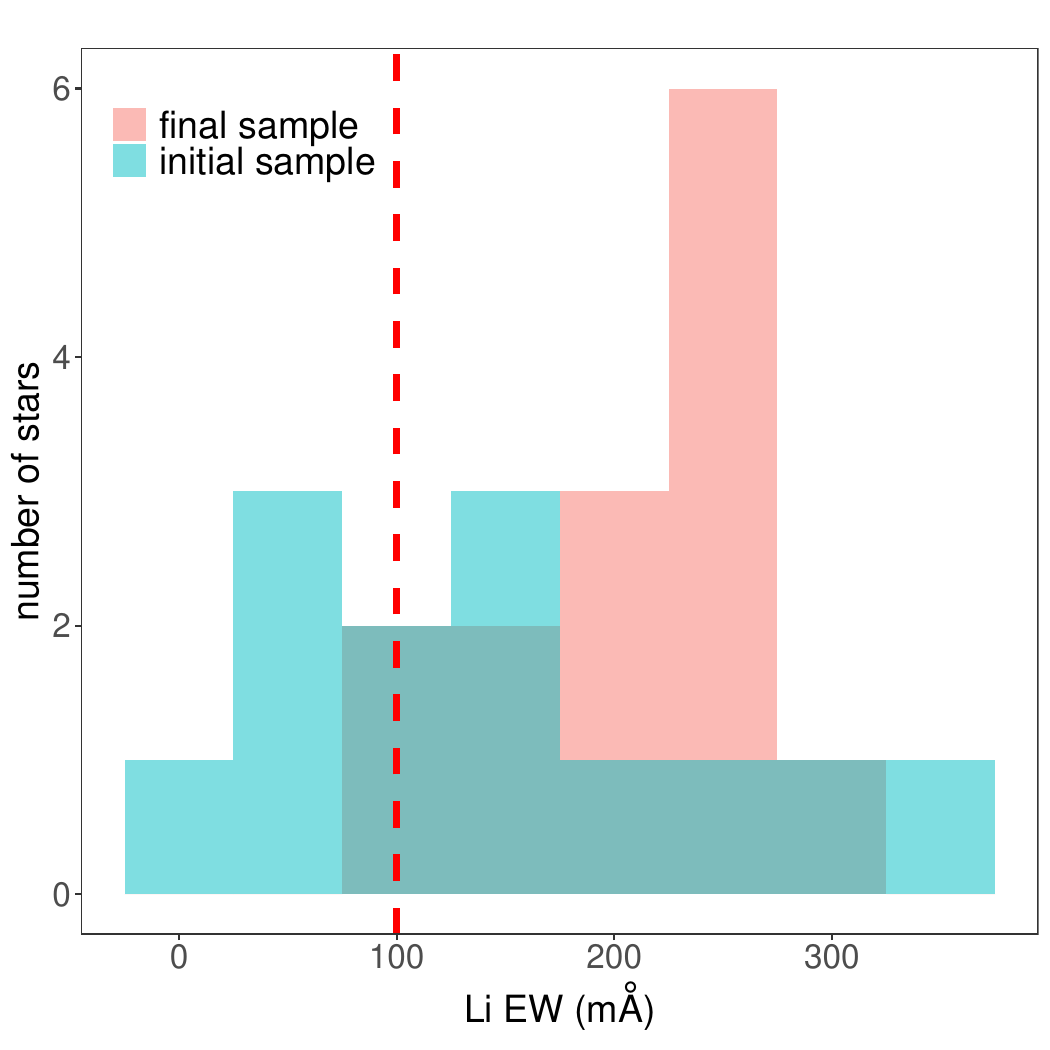}
\caption{
\label{fig_youth_indicators} 
Youth indicators investigated in this study for Octans stars. The red shaded area in the upper-left panel defines the locus in the colour-colour diagram that is representative of young stellar objects with UV emission. The dashed red lines in the histograms denote the adopted thresholds for HR1, MIR excess significance, and lithium EWs used to classify the stars in our sample as likely young.
}
\end{center}
\end{figure*}

First, we inspected the ultraviolet (UV) emission of the stars in our sample, which is a common feature among  young stellar objects. We searched the revised version of the Galaxy Evolution Explorer (GALEX) space telescope catalogue \citep{Bianchi2017} and found near-UV (NUV) magnitudes for 34 stars in our sample. We followed the same procedure employed by \citet{2018ApJ...860...43G}  to select the most likely young stellar objects with UV emission from a colour-colour diagram based on $NUV-G$ and $G-J$ colours. We cross-matched our initial list of stars with the Two Micron All Sky Survey (2MASS) catalogue and retrieved the J-magnitude for 101 stars in the sample. We applied the criterion given by equation~(2) of \citet{2018ApJ...860...43G} that delimitates the region of the colour-colour diagram populated by young stellar objects (see Figure~\ref{fig_youth_indicators}) and identified 21~sources with measured UV emission as young stars.  

Second, we searched the second Röntgen Satellite (ROSAT) all-sky survey catalogue \citep{Boller2016} and the fourth \textit{XMM-Newton} source catalogue \citep[4XMM-DR13;][]{Webb2020} for X-ray emission data and retrieved hardness ratio 1 (HR1) for 31 stars in the sample. We compared the observed distribution of HR1 in young local associations ($\lesssim40$~Myr) and the Hyades open cluster reported by \citet{Kastner2003} and selected all sources in our sample with $\text{HR1}>-0.153$ as likely young. The adopted threshold defines an upper limit within 3$\sigma$ of the mean HR1 value reported for the Hyades in that study. 

Third, we cross-matched our list of stars with the \citet{McDonald2017} catalogue that provides mid-infrared (MIR) excess emission for a number of stars including young stellar objects. The authors provide as part of the catalogue a statistic of the significance of the MIR excess emission ($S_{MIR}$) to help quantify their results. We followed the same strategy employed by \citet{2018ApJ...860...43G} to select the sources with a MIR excess emission significance above 1.5$\sigma$ (i.e. $S_{MIR}\geq1.5$) as young stellar objects. Doing so, we classified 41 stars in our sample as likely young based on this youth indicator. 

Finally, we used the high-resolution spectra collected in this study (see Sect.~\ref{sect2_RV}) to measure the lithium equivalent widths (EWs) of the stars with available spectra in our sample. Lithium absorption (Li $\lambda6708{\AA}$) is one of the primary criteria for stellar youth. We measured the lithium EWs with iSpec by fitting a Gaussian to the line profile after correcting the spectrum for the radial velocity and continuum fitting. For the sources with multiple spectra available for our analysis we provide the median value of the lithium EWs measured from all spectra. Doing so, we measured the lithium EWs of 27 stars in our sample. We considered a lithium EW above 100~m{\AA} to be a strong indication of youth \citep[see also][]{Moor2013,2018ApJ...860...43G} and classify 21 stars in the sample as likely young. 

Doing so, we found at least one sign of youth for 59 stars in our initial list of 103 candidate members of the Octans association. Our final sample of selected stars that will be used for the traceback analysis consists of 29~stars (see  Sections~\ref{sect2_sample_selection} and \ref{section3}) and we found youth indicators for 20 of them. Table~\ref{tabA3} lists all the data for the youth indicators investigated in our analysis. Two points are worth mentioning in regard to the youth diagnostics performed in this study. First, we used fixed thresholds (i.e. hard cuts in the sample) for the different youth indicators to identify the most likely young stellar objects. The stars that were not flagged as likely young in our analysis are not necessarily inconsistent with youth. The locus of each object class (`young' vs. `not young') defines a much more complex shape in the space of youth  parameters that requires further investigation with a much more significant sample of young stars, but this clearly goes beyond the scope of this paper. Second, it is important to mention that many stars in our sample are still lacking data for the youth indicators explored in our analysis. This makes our classification scheme for young stellar objects in the Octans association rather incomplete.

\subsection{Sample selection}\label{sect2_sample_selection}

As explained in Sect~\ref{sect2_RV} the radial velocities derived in this paper are mostly more precise than the ones given in the \textit{Gaia}-DR3 catalogue, but they appear in small number. We obtained the largest sample of Octans stars with radial velocity information by combining our measurements with the \textit{Gaia}-DR3 catalogue. For the sources studied in both projects, we took the most precise radial velocity measurement that is available. Doing so, we find a sample of 86~stars with complete 6D data (2D positions, proper motions, parallaxes and radial velocities). 

\begin{figure*}
\begin{center}
\includegraphics[width=0.33\textwidth]{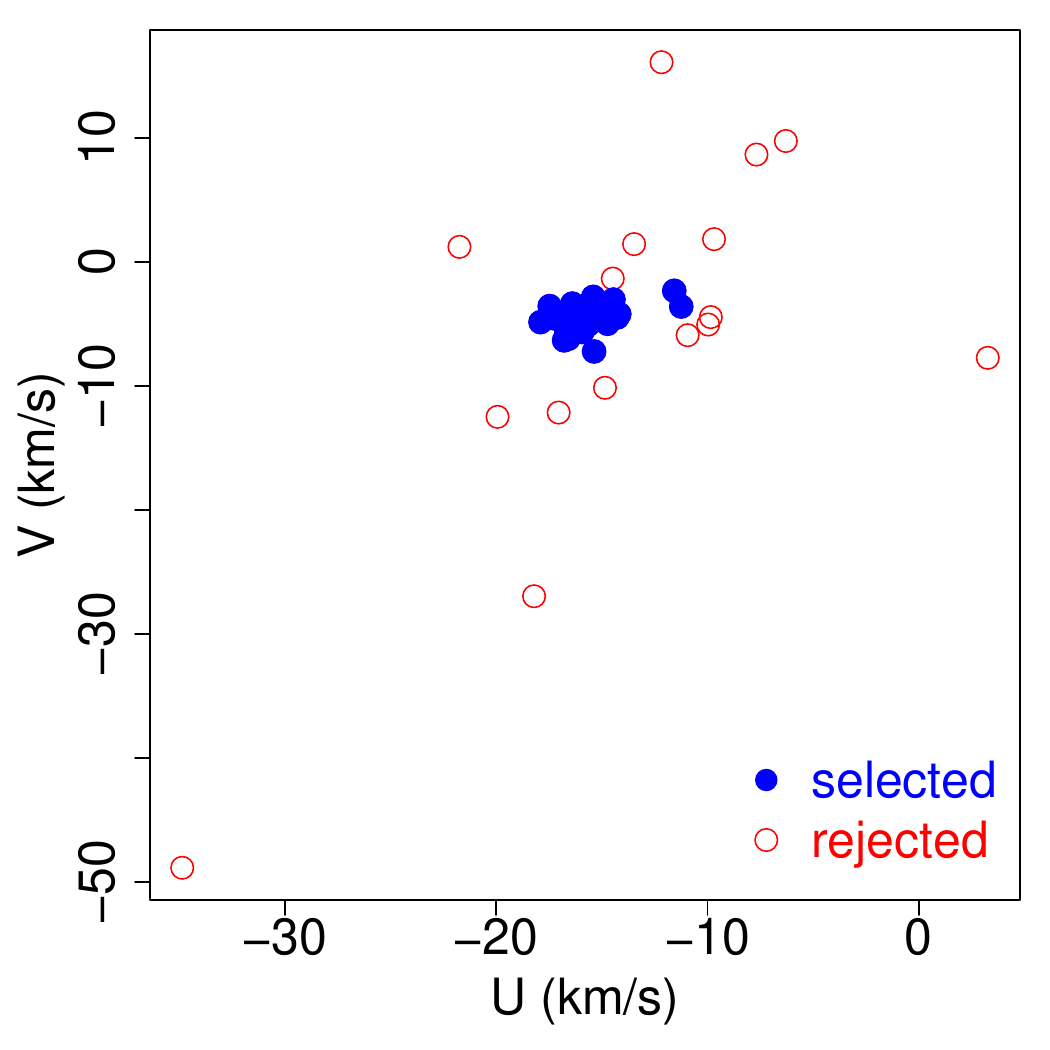}
\includegraphics[width=0.33\textwidth]{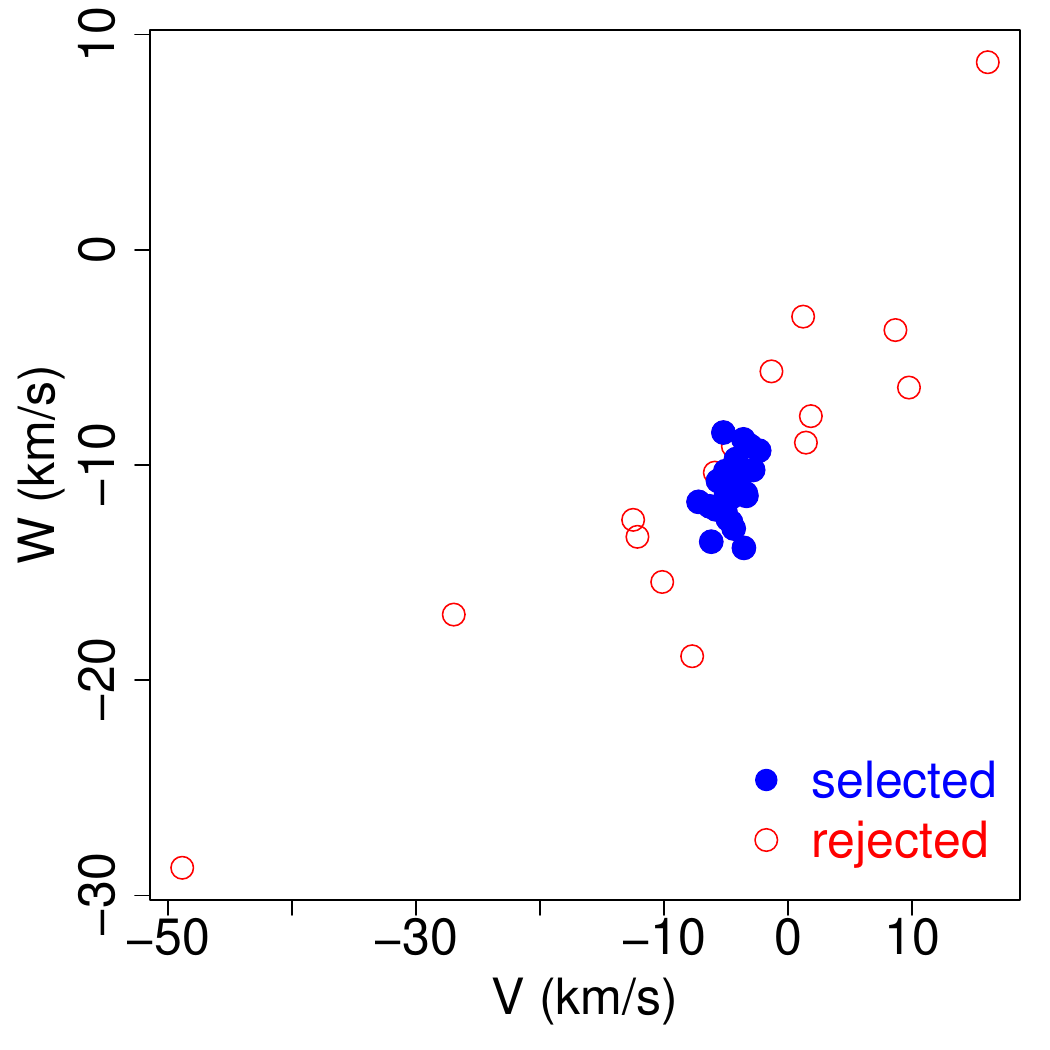}
\includegraphics[width=0.33\textwidth]{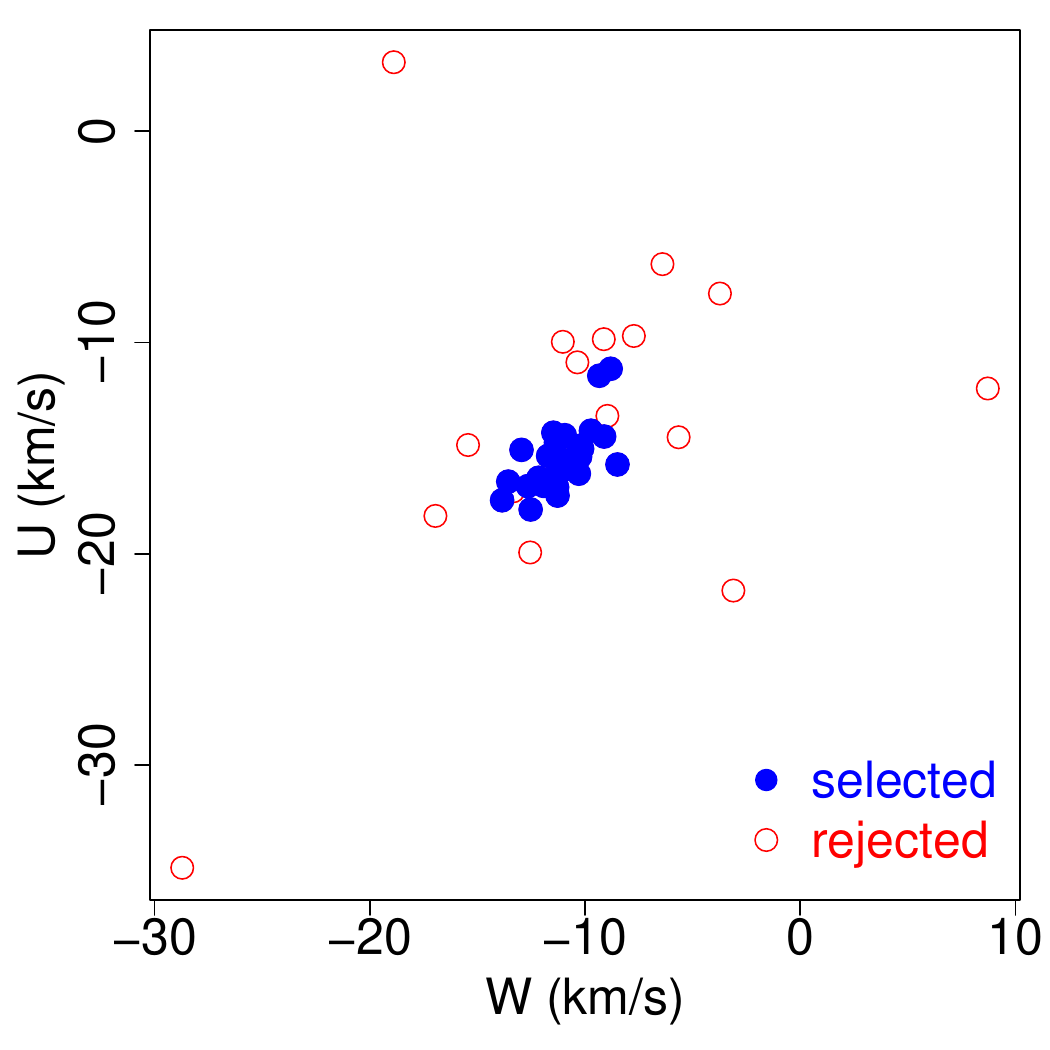}
\caption{
\label{fig_clean_UVW_sample} 
3D spatial velocity of Octans stars and sample selection obtained from the MCD method.
}
\end{center}
\end{figure*}

As discussed in previous studies conducted by our team \citep{2020A&A...642A.179M,2023MNRAS.520.6245G} the results obtained for the dynamical age are sensitive to the quality of the data and the existence of outliers in the sample. It is therefore important to clean the sample by removing the sources with poor measurements and to select the most likely kinematic members of the association before running the traceback method. To do so, we first selected the sources with radial velocity precision better than 1~km/s and discarded known binaries in the sample due to the variable radial velocity of these sources. This yields a sample of 53~stars. Then, we computed the 3D space motion of the stars and removed potential outliers in the 3D space of velocities. The $UVW$ Galactic velocity of the stars was computed in the same reference system defined by \citet{1987AJ.....93..864J}, where $X$ points to the Galactic centre, $Y$ points in the direction of Galactic rotation, and $Z$ points to the Galactic north pole. We computed robust distances (RDs) to select the most likely members of the association based on their 3D space motion:
\begin{equation}
RD(\mathbf{x})=\sqrt{(\mathbf{x}-\boldsymbol{\mu})^{t}\,\boldsymbol{\Sigma}^{-1}(\mathbf{x}-\boldsymbol{\mu})}
\end{equation}
where $\boldsymbol{\mu}$ and $\boldsymbol{\Sigma}$ denote the multivariate location and covariance matrix of the data that are obtained from the minimum covariance determinant \citep[MCD;][]{Rousseeuw1999} method. We used a tolerance ellipse of 99\% to select the most likely kinematic members of the association and remove potential outliers from the sample. This procedure retains 31 stars in the sample and the results of this analysis are illustrated in Figure~\ref{fig_clean_UVW_sample}.

One last step of the sample selection consists in inspecting the convergence of the stellar orbits. The members of a stellar association share a common origin, so one would expect the stellar orbits to converge in the past. Thus, the common approach of selecting cluster members based only on the present-day position and velocity of the stars will fail to remove potential interlopers with diverging orbits in the past \citep[see][]{2023MNRAS.520.6245G}. After removing the sources with diverging orbits (see also Sect.~\ref{section3}), we ended up with an effective sample of 29~stars that can be used in the traceback analysis. 

The mean space motion that we find for the Octans stellar association based on our final sample of cluster members is $(U,V,W)=(-15.7,-4.6,-11.1)\pm(0.2,0.2,0.2)$~km/s. The 1D velocity dispersion in each direction obtained from the standard deviation of the velocities is $(\sigma_{U},\sigma_{V},\sigma_{W})=(1.2,1.0,1.2)$~km/s while the median uncertainty in each velocity component is 0.1, 0.4, and 0.2, respectively. This implies that Octans is moving with a low intrinsic velocity dispersion (ranging from 0.8 to 1.2~km/s in the velocity components) as one would expect for a young stellar association. 

We took advantage of this clean sample of cluster members to revisit the distance to the Octans association based on \textit{Gaia}-DR3 parallaxes. We used the Kalkayotl code \citep{2020A&A...644A...7O} that implements a number of priors to compute distances from Bayesian inference. Using a Gaussian prior we find the distance of $151\pm4$~pc confirming that Octans is indeed one of the most distant young stellar associations of the solar neighbourhood. This puts the Octans association somewhat farther away as compared to the previous distance estimate of $130^{+30}_{-20}$~pc given by \citet{2018ApJ...856...23G}.

We repeated the analysis described throughout the manuscript with different samples of Octans stars to ensure the robustness of our results for the dynamical age of the association. To do so, we defined two control samples (CSs), one using only the radial velocities we derived (CS1 hereafter) and the other using only the \textit{Gaia}-DR3 radial velocities (CS2 hereafter). We present in Table~\ref{tab_samples} the number of stars retained in the different samples investigated in this study after each step of the sample selection procedure described above. 

\begin{table*}
\centering
\caption{Number of stars retained in each sample after applying our selection criteria.}
\label{tab_samples}
\begin{tabular}{lccc}
\hline\hline
&CS1&CS2&Sample\\
\hline\hline
Stars with radial velocity data & 34 stars & 86 stars & 86 stars\\
Sample with precise radial velocity data ($\sigma_{RV}<1$~km/s)&26 stars& 52 stars& 61 stars\\
Sample of single stars with 6D data&23 stars& 45 stars& 53 stars\\
Sample of selected single stars in the velocity space&15 stars& 30 stars& 31 stars\\
Final sample confirmed by orbital analysis&11 stars& 25 stars& 29 stars\\
\hline\hline
\end{tabular}
\end{table*}

\section{Traceback}\label{section3}

The traceback analysis used in this paper to infer the dynamical age of the Octans association is based on the same strategy employed in previous studies conducted by our team for other young stellar groups \citep{2020A&A...642A.179M,2023MNRAS.520.6245G}. The main steps of our methodology are summarized below. 

\begin{figure*}
\begin{center}
\includegraphics[width=0.99\textwidth]{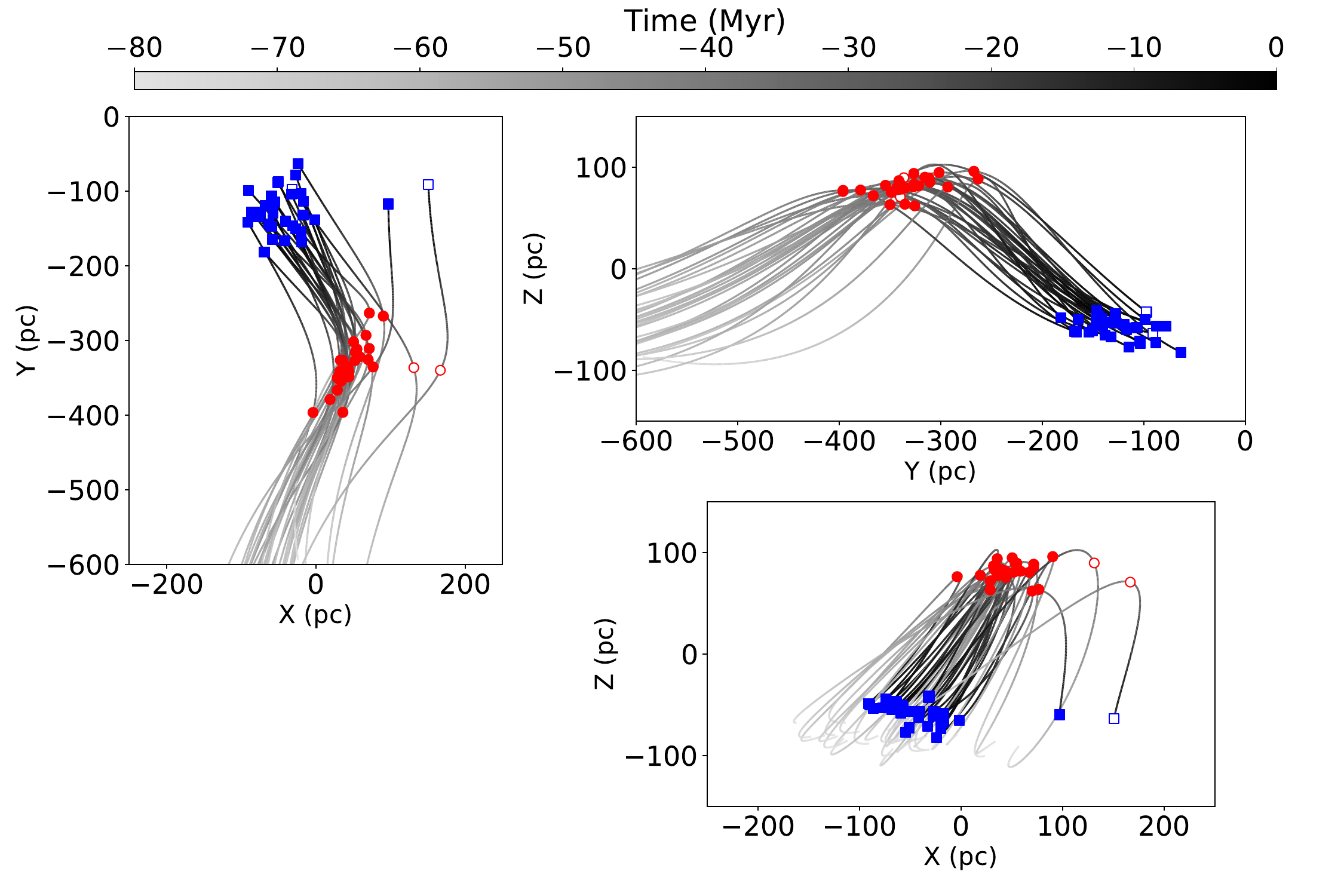}
\caption{
\label{fig_2d_orbits} 
2D projection of the stellar orbits integrated back in time. Blue squares and red circles indicate the present-day position of the stars $(t=0)$ and location at birth time ($t=-34$~Myr), respectively. Open symbols denote the two stars with diverging orbits that have been removed from our sample (see Section~\ref{section3}).
}
\end{center}
\end{figure*}

The stellar orbits are integrated in the past using the Milky Way's galactic potential (hereafter, \texttt{MWPotential2014}) of the \textit{galpy} python library implemented by \citet{2015ApJS..216...29B}. We computed the stellar positions back in time from $t=0$ (present-day position) to $t=-80$~Myr in steps of 0.1~Myr. We used the solar motion of $(U_{\odot},V_{\odot},W_{\odot})=(11.10,12.24,7.25)$~km/s reported by \citet{2010MNRAS.403.1829S} in our analysis. As illustrated in Figure~\ref{fig_2d_orbits} there are two sources (\textit{Gaia} DR3 5568830136157337728 and \textit{Gaia} DR3 6629609336546072320) with orbits that diverge in the past from the remaining stars of the group despite their consistent present-day location and velocity. We manually removed these stars from the sample, as explained in Sect.~\ref{sect2_sample_selection}. 

We investigated different metrics to define the size of the association, which is needed to infer the dynamical age. First, we estimated the size of the association based on the dispersion of the stellar positions in the $X$,$Y$, and $Z$ directions (hereafter, $S_{X}$,$S_{Y}$, and $S_{Z}$) from the square-root of the elements in the main diagonal of the covariance matrix. Second, we used the trace of the covariance matrix to define the size of the association as
\begin{equation}
S_{TCM}=\left[ \frac{\mathrm{Tr}(\Sigma)}{3} \right]^{1/2}\,,
\end{equation}
where $\Sigma$ is the covariance matrix computed from the MCD method (see Sect.~\ref{sect2_sample_selection}). The trace of the covariance returns the total variance of the data and the multiplicative factor of $1/3$ included in the definition of $S_{TCM}$ returns the arithmetic mean of the variances in all three directions. Finally, we also estimated the size of the association from the determinant of the covariance matrix as
\begin{equation}
S_{DCM}=[\det(\Sigma)]^{1/6}\,,
\end{equation}
which roughly defines the volume of the association. In the following we present our results for the dynamical age of the association obtained from these metrics and discuss the differences. 

\begin{figure}
\begin{center}
\includegraphics[width=0.49\textwidth]{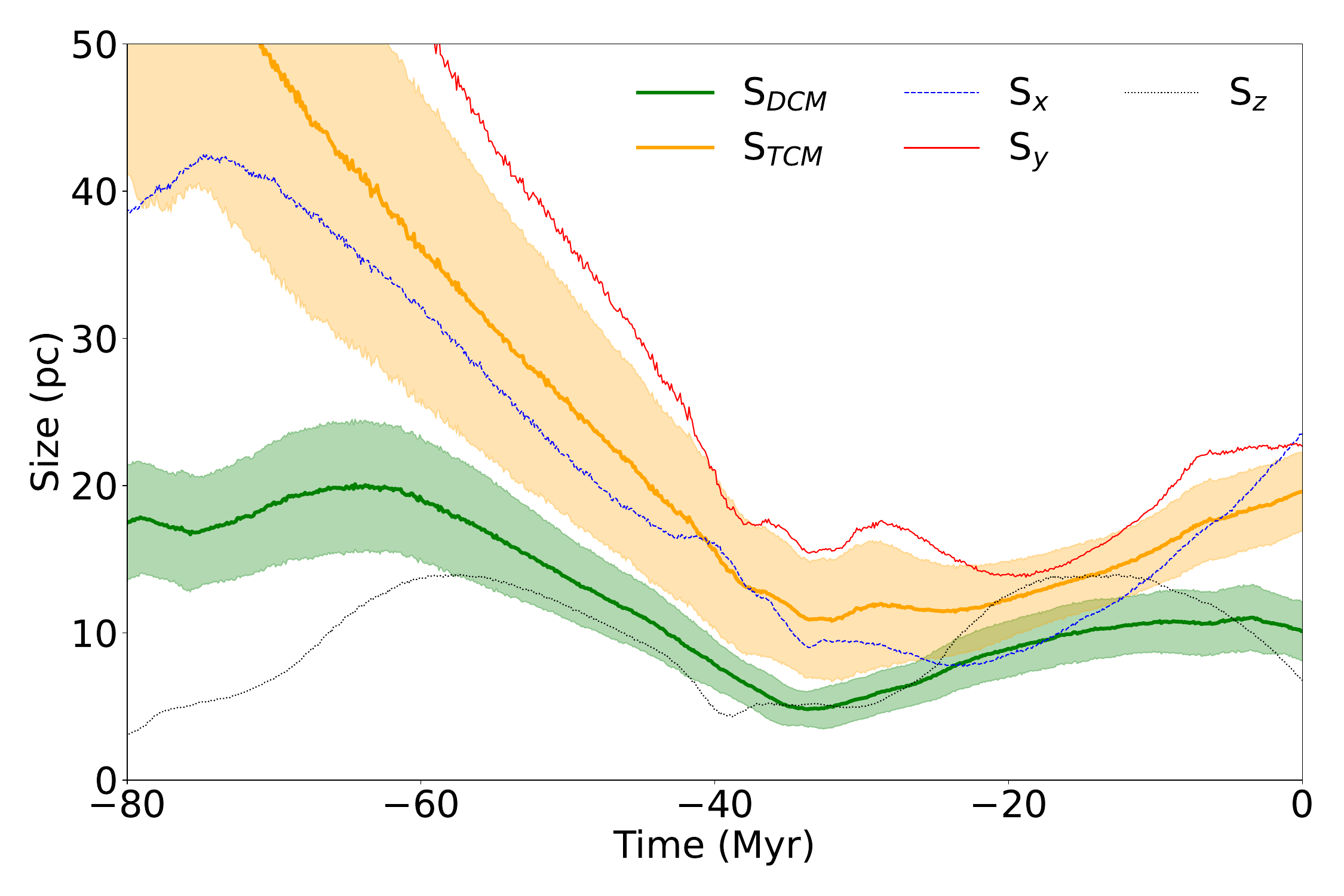}
\caption{
\label{fig_age_size} 
Size of the association as a function of time for the different metrics employed to investigate the size of the association in this study. The shaded regions denote the $1\sigma$ uncertainties in the size of the association obtained from 1\,000 bootstrap repetitions. The lines indicate the median values for the dynamical age obtained from each size estimator of the association. 
}
\end{center}
\end{figure}

Figure~\ref{fig_age_size} shows the size of the association computed as a function of time for the different metrics. It is apparent that the $S_{X}$, $S_{Y}$ and $S_{Z}$ size estimators produce different results with dynamical ages ranging from 24 to 39~Myr that are clearly not consistent among themselves. On the other hand, we note that the results obtained from $S_{TCM}$ and $S_{DCM}$ are in good agreement and the minimum of the curves occurs at about 33~Myr. However, it is important to mention that the observed uncertainties in the size of the association computed from the $S_{TCM}$ size estimator are larger, which makes the locus of the minimum of the curve and the dynamical age more uncertain. We therefore decided to report the more robust solution obtained from $S_{DCM}$ as our final result for the dynamical age of the association. 

\begin{figure}[!h]
\begin{center}
\includegraphics[width=0.48\textwidth]{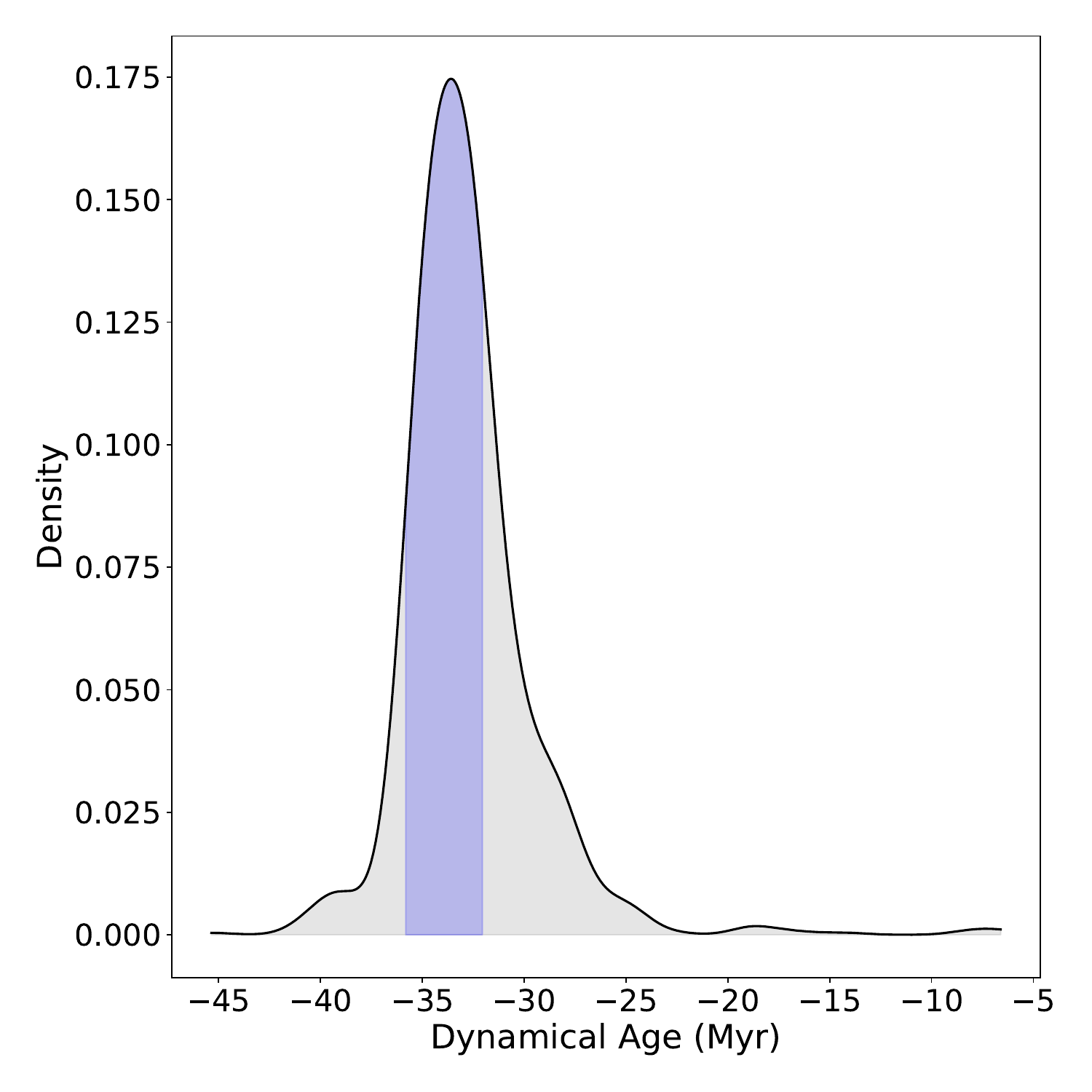}
\caption{Distribution of dynamical ages obtained after 1 000 bootstrap repetitions for our sample of 29 stars using the $S_{DCM}$ size estimator and the \texttt{MWPotential2014} in the traceback analysis. The blue shaded region indicates the 68\% highest-density interval of our solution. \label{fig_age_bootstrap} 
}
\end{center}
\end{figure}

Table~\ref{tab_age} lists our results obtained for the different samples of stars and size estimators of the association investigated in this paper. The dynamical ages given in Table~\ref{tab_age} for each solution result from a bootstrap procedure as explained below. First, we performed 1\,000 bootstrap repetitions using random samples of cluster members and compute the dynamical age of each bootstrap sample from the traceback analysis as described before. Then, we took the mode and the 68\% highest-density interval of the ensemble of ages from the bootstrap samples to define the dynamical age and its uncertainty for each solution. We note that the dynamical ages inferred from the $S_{DCM}$ size estimator for the three samples are consistent among themselves and agree within 1~Myr. In particular, we note that the dynamical age obtained from our final sample is more precise than the results obtained with CS1 and CS2 thanks to the higher number of stars employed in the traceback analysis. Figure~\ref{fig_age_bootstrap} illustrates the distribution of dynamical age obtained from the $S_{DCM}$ size estimator. The bootstrap repetitions obtained in our analysis reveal a somewhat skewed distribution of dynamical ages with minor impact in the uncertainty of our solution.

\begin{table}
\centering
\caption{Dynamical age of Octans computed from different size estimators of the association employed in the traceback analysis.}
\label{tab_age}
\begin{tabular}{lcccccc}
\hline\hline
&Stars&$S_{X}$ & $S_{Y}$ & $S_{Z}$ & $S_{TCM}$ & $S_{DCM}$\\
&&(Myr)&(Myr)&(Myr)&(Myr)&(Myr)\\
\hline\hline
\vspace{0.1cm}
CS1&11& $28^{+5}_{-3}$ & $23^{+9}_{-3}$ & $36^{+3}_{-6}$ & $30^{+4}_{-3}$ & $35^{+2}_{-9}$ \\
CS2&25& $24^{+4}_{-7}$ & $16^{+13}_{-4}$ & $40^{+1}_{-4}$ & $29^{+5}_{-9}$ & $34^{+2}_{-5}$ \\
Sample&29& $24^{+10}_{-2}$ & $25^{+12}_{-4}$ & $39^{+2}_{-5}$ & $33^{+2}_{-7}$ & $34^{+2}_{-2}$ \\
\hline\hline
\end{tabular}
\end{table}

\begin{table}
\centering
\caption{Results for the dynamical age of Octans using different models of the Galactic potential to integrate the stellar orbits.}
\label{tab_age_potential}
\begin{tabular}{lccccc}
\hline\hline
Model& $S_{X}$ & $S_{Y}$ & $S_{Z}$ & $S_{TCM}$ & $S_{DCM}$\\
&(Myr)&(Myr)&(Myr)&(Myr)&(Myr)\\
\hline\hline
\texttt{MWPotential2014}& $24^{+10}_{-2}$ & $25^{+12}_{-4}$ & $39^{+2}_{-5}$ & $33^{+2}_{-7}$ & $34^{+2}_{-2}$ \\
\texttt{Irrgang13I}& $22^{+11}_{-1}$ & $24^{+10}_{-6}$ & $39^{+2}_{-5}$ & $30^{+4}_{-5}$ & $32^{+3}_{-1}$  \\
\texttt{McMillan17}& $22^{+9}_{-2}$ & $23^{+13}_{-3}$ & $38^{+2}_{-7}$ & $32^{+2}_{-7}$ & $34^{+1}_{-3}$  \\

\hline\hline
\end{tabular}
\end{table}

The \textit{galpy} package includes a variety of potentials that can be used to integrate orbits. We took advantage of this facility and repeated the traceback analysis using different models for the Galactic potential. In addition to the \texttt{MWPotential2014} potential, we also performed our calculations for the orbit integration with the \texttt{Irrgang13I} \citep[model~I from][]{2013A&A...549A.137I} and \texttt{McMillan17} \citep{2017MNRAS.465...76M} potentials. The \texttt{Irrgang13I} model includes a \citet{1975PASJ...27..533M} potential for bulge and axisymmetric disc, and a modified dark matter halo potential from \citet{1991RMxAA..22..255A}. The \texttt{McMillan17} potential consists of an axisymmetric approximation of the \citet{2002MNRAS.330..591B} bulge model, thin and thick disc with exponentially decreasing density profile, and a dark matter halo with the \citet{1996ApJ...462..563N} density profile. In Table~\ref{tab_age_potential} we compare our results for the dynamical age of the association using different potentials. We note that the difference in the dynamical ages computed from different size estimators and samples of stars is mostly smaller than 1-2~Myr when we change the galactic potential model that is used in the traceback analysis. The observed difference is smaller than the typical age uncertainties reported in our results. We therefore confirm that our result for the dynamical age of the Octans association does not depend on the choice of the Galactic potential. A similar conclusion was reached in previous studies that targeted other young stellar groups \citep[see e.g.][]{2020A&A...642A.179M,2023MNRAS.520.6245G}.

In a recent study, \citet{2023ApJ...946....6C} investigated the impact of several sources of systematic errors on the computation of traceback ages. In particular, they argue that uncorrected gravitational redshift and convective blueshift produce a bias of about 0.6~km/s in the radial velocity measurements, which results in younger traceback ages by about 2~Myr for the specific case of the $\beta$~Pictoris moving group. We investigated the impact of this correction for the case of the Octans association as explained below. The spectral type of the stars in our final sample that is used in the traceback analysis ranges from about A1 to G9. As illustrated in Figure~3 of \citet{2023ApJ...946....6C} the total radial velocity shift (including the effects of gravitational redshift and convective blueshift) is expected to be about 0.6~km/s for the stars in our sample \citep[see also][]{Gagne2024}. After correcting the radial velocities and recomputing the traceback analysis, we find a dynamical age of $33^{+3}_{-1}$~Myr, which is fully consistent with our previous solution within the reported uncertainties. We therefore conclude that the corrections due to gravitational redshift and convective blueshift have no significant impact on our age estimate obtained for the Octans association.

\nopagebreak
\section{Discussion}\label{section4}

\begin{figure*}[!h]
\sidecaption
\includegraphics[width=6cm]{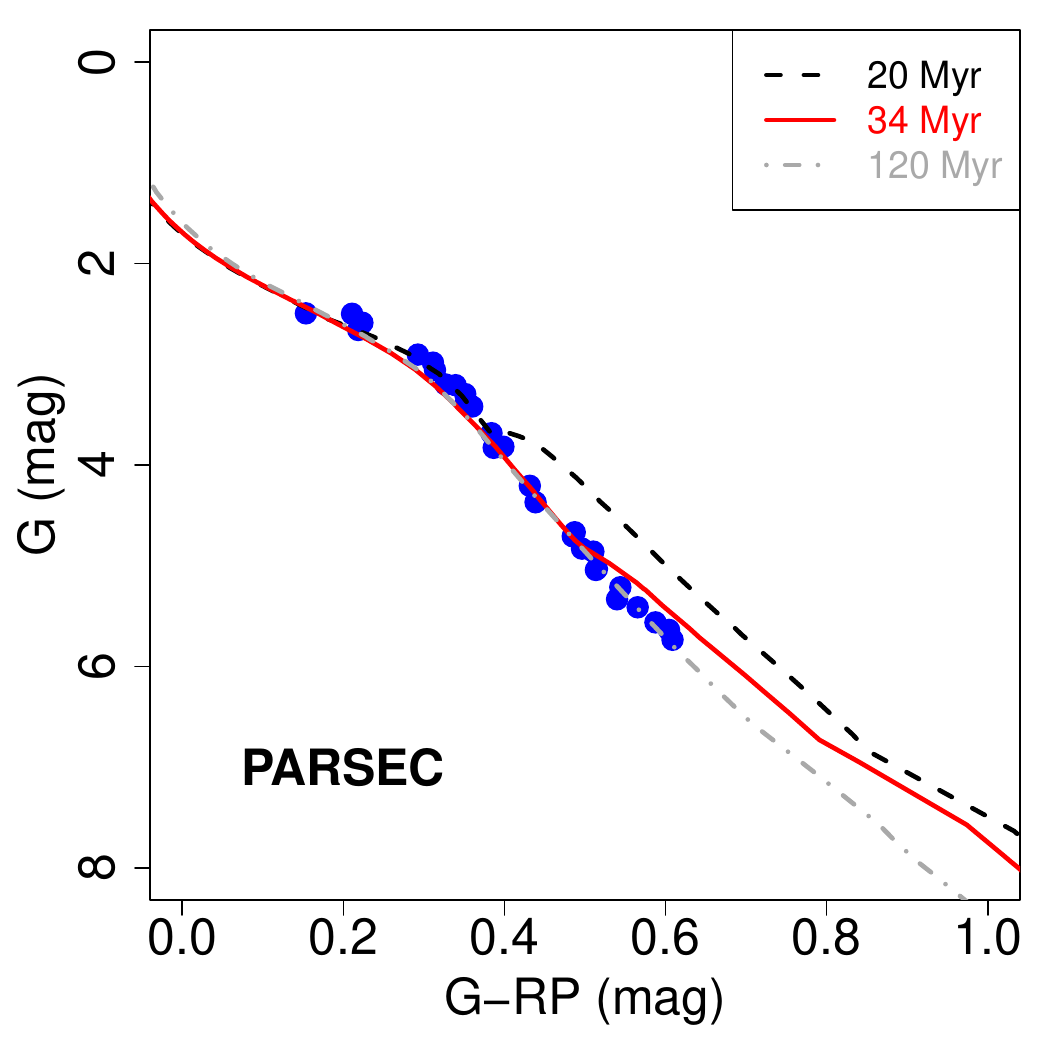}
\includegraphics[width=6cm]{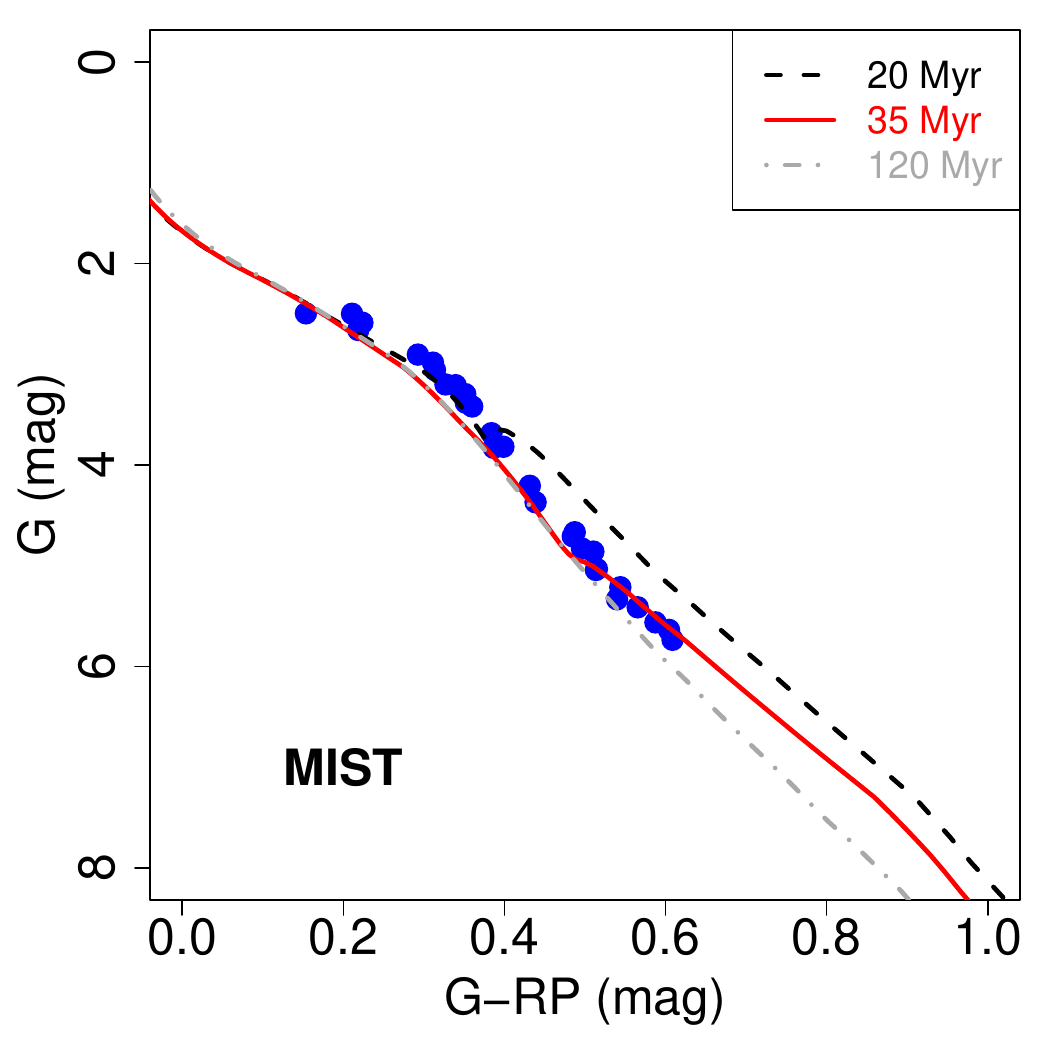}
\caption{Absolute colour magnitude diagram of Octans stars based on the \textit{Gaia}-DR3 photometry. The dashed, solid, and dotted lines indicate the 20~Myr, 34~Myr, and 120~Myr isochrones from PARSEC v1.2S \textit{(left panel)} and MIST with $v/v_{crit}=0.4$ \textit{(right panel)} models, respectively.
\label{fig_CMD} 
}
\end{figure*}

Early studies after recognition of Octans stars as a young stellar association assigned the age of 20-30~Myr based on isochrones \citep{Torres2003a,Torres2003b,Torres2008}. As illustrated in Figure~\ref{fig_CMD}, the lower limit of 20~Myr reported in the literature clearly underestimates the observed age of the association based on our sample of selected cluster members (see Sect.~\ref{sect2_sample_selection}). From a visual inspection of Figure~\ref{fig_CMD} we note that most cluster members are distributed along the 34~Myr isochrone computed from the PARSEC v1.2S \citep{PARSEC} and MIST \citep{MIST} models that cover the entire mass range of our sample. On the other hand, we confirm that the dynamical age we obtained supports the LDB age estimate of 30-40~Myr given by \citet{Murphy2015} and provides a more robust (and precise) age estimate for the association. To the best of our knowledge the result presented in this paper is the first reliable dynamical age estimate for the Octans association that is currently available. The previous study conducted by \citet{MiretRoig2018} performed a traceback analysis of the Octans association using data from the \textit{Tycho-Gaia} astrometric solution \citep{GaiaDR1}, but they found an ambiguous result that only allowed them to set a lower limit of 4~Myr for the dynamical age of the association. Figure~\ref{fig_CMD} also illustrates the differences that exist among different evolutionary models confirming the importance of deriving stellar ages that are independent from stellar models.

The dynamical age obtained in this paper makes Octans older than the $\beta$~Pictoris moving group \citep[$18.5^{+2.0}_{-2.4}$~Myr;][]{2020A&A...642A.179M} and coeval with the Tucana-Horologium association \citep[$38.5^{+1.6}_{-8.0}$~Myr;][]{2023MNRAS.520.6245G} within the reported uncertainties. The dynamical ages obtained for these young stellar associations were computed from the same traceback methodology, which allows for a direct comparison of the results. These young stellar associations are believed to belong to different systems of plausibly related nearby young stellar groups \citep{2021ApJ...915L..29G}, but computing their ages with the same methodology is the first step towards constructing a consistent age-scale for the young clusters of the solar neighbourhood.  

The minimum size of the Octans association derived from the $S_{DCM}$ size estimator is about 5~pc. This is consistent with the minimum size derived for the $\beta$~Pictoris moving group \citep[$\sim7$~pc;][]{2020A&A...642A.179M} and the Tucana-Horologium association \citep[$\sim4$~pc;][]{2023MNRAS.520.6245G}. The size of the Octans association in the $XYZ$ directions inferred from the $S_{X}$,$S_{Y}$, and $S_{Z}$ estimators at birth time ($t=-34$~Myr) is 9.1, 15.5, and 5.1~pc, respectively. This also confirms that the typical size of a stellar association at birth is a few parsecs. 

We provide in Table~\ref{tabA4} the 3D~positions ($XYZ$) and 3D velocities ($UVW$) of the association at present time and birth time. The 1D velocity dispersion of the stars (i.e. standard deviation) at $t=-34$~Myr is $\sigma_{U}=1.1$~km/s, $\sigma_{V}=1.2$~km/s, and $\sigma_{W}=1.2$~km/s. It is only somewhat smaller than the observed present-day velocity dispersion ($\sigma_{U}=1.2$~km/s, $\sigma_{V}=1.0$~km/s, $\sigma_{W}=1.2$~km/s; see Sect.~\ref{sect2_sample_selection}), which indicates that the motion inherited from the molecular clouds is well preserved over a time interval of  $\sim30$~Myr. This low velocity dispersion is expected for a young stellar association and is also consistent with the reported velocity dispersion in nearby star-forming regions \citep[see e.g.][]{2019A&A...630A.137G,2023A&A...671A...1O}.

It is also worth mentioning that the dynamical age for Octans derived in this study has the same level of precision of the result from \citet{2020A&A...642A.179M} for the $\beta$~Pictoris moving group, which is located at a distance of $\sim40$~pc. Previous applications of the traceback technique have mostly focused on the closest young stellar associations \citep[see e.g.][]{2006AJ....131.2609D,2014A&A...563A.121D,2023MNRAS.520.6245G}. Our result demonstrates the feasibility of deriving reliable and precise dynamical ages for $\sim30$~Myr old young stellar associations at about $\sim150$~pc based on the traceback technique. 

One intriguing question that arises is the relationship of the Octans association with other nearby stellar groups in the sky. \citet{2013ApJ...778....5Z} identified a nearby ($<100$~pc) group of co-moving stars with the Octans association with apparent ages between 30 and 100~Myr that they proposed as a new association named Octans-Near. \citet{Murphy2015} argued that the two associations could belong to a larger complex that has undergone several star formation episodes over the past 100~Myr that would explain the large age range in Octans-Near. Other subsequent studies after the discovery of Octans-Near as a potential new group of the solar neighbourhood argued that it is likely composed of non-coeval stars with a wide range of velocities \citep{2016IAUS..314...21M,2017AJ....153...95R}. In the following we investigate a possible connection between Octans and Octans-Near using the traceback method. 

Our initial sample of stars for this analysis includes the 21~stars listed in Table~1 of \citet{2013ApJ...778....5Z} as probable and possible members of the Octans-Near association. We cross-matched this list of stars with the \textit{Gaia}-DR3 catalogue and retrieved the 5D astrometry for 20 stars and radial velocities for 17~stars. We discarded binaries and multiple systems from the sample to avoid biased results in the traceback analysis from sources with variable radial velocity. This yields an effective sample of 10~stars with complete data (5D astrometry and radial velocity) that can be used in our analysis.  

\begin{figure*}[!h]
\begin{center}
\includegraphics[width=0.9\textwidth]{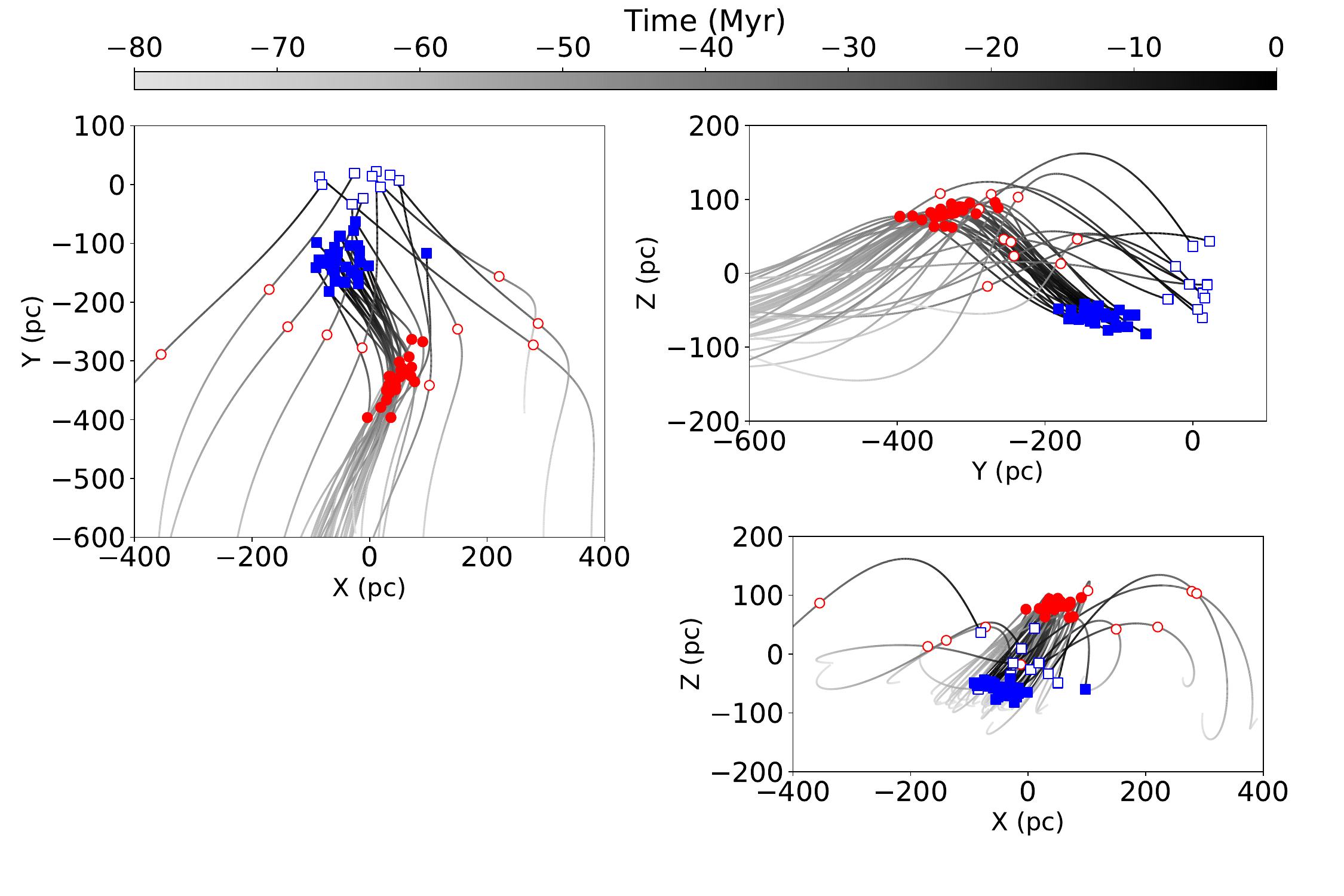}
\caption{2D projection of the stellar orbits integrated back in time for Octans and Octans-Near stars. Filled symbols and open symbols indicate the members of the Octans and Octans-Near associations, respectively. Blue squares denote the present-day location of the stars ($t=0$), and red circles mark the stellar positions at the birth time of the Octans association ($t=-34$~Myr).
\label{fig_orbits_OCTN} }
\end{center}
\end{figure*}

We integrated the stellar orbits in the past using the \texttt{MWPotential2014} galactic potential and compared the stellar positions of the two stellar associations at $t=-34$~Myr, which roughly corresponds to the birth time of the Octans association. The results of this analysis are illustrated in Figure~\ref{fig_orbits_OCTN}. It is apparent that Octans and Octans-Near stars were clearly located at different positions in the past indicating that they formed at different locations. The different location in the past combined with the different age and space motion of the stars (as reported in previous studies) suggests that Octans and Octans-Near are more likely to be unrelated. Moreover, it is also interesting to note that the stellar orbits of Octans-Near stars appear to diverge in the past instead to clump in space to define a configuration of minimum size as we observe for Octans itself. This raises questions about the integrity of the association and suggests that Octans-Near is possibly an artefact (not a stellar association). 

\section{Conclusions}\label{section5}

In this work we investigated the dynamical age of the Octans young stellar association using the precise astrometric data and radial velocities given by the \textit{Gaia}-DR3 catalogue. We complemented the \textit{Gaia}-DR3 radial velocities with additional measurements for 34~stars in our sample obtained from observations we performed with the CHIRON spectrograph and high-resolution spectra downloaded from public archives. The radial velocities derived in this study are consistent with the \textit{Gaia}-DR3 results, and they are mostly more precise for the sources in common. 

As part of our strategy for deriving reliable age estimates, we performed a careful selection of the stars in our sample before running the traceback analysis. We discarded binaries and sources with poor radial velocity information from the sample, rejected potential outliers in the 3D space of velocities, and inspected the convergence of the stellar orbits back in time. Then, we performed an extensive analysis of the results obtained using the traceback method. We recomputed the dynamical age with different samples of Octans stars, explored different metrics for evaluating the size of the association, and tested different models for the Galactic potential to integrate the stellar orbits. The dynamical ages computed from  different samples of stars and models for the Galactic potential show little variation and are consistent within the reported uncertainties, which confirms the robustness of our solution. 

We report the dynamical age of $34^{+2}_{-2}$~Myr, which is the first reliable dynamical age obtained for Octans. After correcting the radial velocities of the stars for gravitational redshift, we find a dynamical age of $33^{+3}_{-1}$~Myr. The small difference between the two solutions suggests that the effect of gravitational redshift is not significant here. Our result is consistent with the less accurate age estimates of 30-40~Myr given by the LDB method and 20-30~Myr from isochrones that are published in the literature. The dynamical age obtained in this study is therefore the most precise and robust age estimate currently available for the Octans association. Our results also demonstrate the feasibility of deriving reliable dynamical ages for $\sim30$~Myr old stellar groups at about $\sim150$~pc with the same level of precision that has been achieved for the nearby $\beta$~Pictoris moving group \citep[dynamical age of $18.5^{+2.0}_{-2.4}$~Myr;][]{2020A&A...642A.179M} located at about $\sim40$~pc.

Finally, by integrating the orbits of the stars in Octans and Octans-Near, we show that the two associations were located at different positions in the past. Previous studies argued that Octans-Near is composed of non-coeval stars with a wide range of velocities, and our traceback analysis supports the scenario of a different origin for the stars in these two associations. It is also apparent from our traceback analysis that the stars in Octans-Near diverged from each other in the past, raising questions about the integrity of this stellar group.

\begin{acknowledgements}
We thank the referee for constructive criticism that helped us to improve the manuscript. P.A.B.G. acknowledges financial support from the São Paulo Research Foundation (FAPESP) under grants 2020/12518-8 and 2021/11778-9. J.O. acknowledges financial support from ``Ayudas para contratos postdoctorales de investigación UNED 2021''. D.B. has been funded by Spanish MCIN/AEI/10.13039/501100011033 grant PID2019-107061GB-C61. This research has received funding from the European Research Council (ERC) under the European Union’s Horizon 2020 research and innovation programme (grant agreement No 682903, P.I. H. Bouy), and from the French State in the framework of the  ``Investments for the future” Program, IdEx Bordeaux, reference ANR-10-IDEX-03-02. This research has made use of the SIMBAD database, operated at CDS, Strasbourg, France. This work has made use of data from the European Space Agency (ESA) mission {\it Gaia} (\url{https://www.cosmos.esa.int/gaia}), processed by the {\it Gaia} Data Processing and Analysis Consortium (DPAC, \url{https://www.cosmos.esa.int/web/gaia/dpac/consortium}). Funding for the DPAC has been provided by national institutions, in particular the institutions participating in the {\it Gaia} Multilateral Agreement. 
\end{acknowledgements}

\bibliographystyle{aa} 
\bibliography{references} 

\onecolumn
\begin{appendix} 
\section{Additional tables}

\begin{table*}[!h]
\centering
\caption{Radial velocities derived in this paper for Octans stars. This table will be available in its entirety at the CDS.
\label{tabA1}}
\resizebox{18cm}{!}{%
\begin{tabular}{lcccccccc}
\hline\hline
Source identifier&$\alpha$&$\delta$&MJD&$RV$&$\sigma_{RV}$&Program ID&Instrument&Flag\\
&(deg)&(deg)&(days)&(km/s)&(km/s)&&&\\
\hline\hline

4836484576636946560 & 57.62286029 & -43.6182879 & 59436.4 & $ 14.75 \pm 0.09 $& 0.24 & 661 & CHIRON & 0 \\
4823999931341641600 & 77.25054007 & -36.4610608 & 58746.3 & $ 15.71 \pm 0.05 $& 0.17 & 0103.C-0759 & HARPS & 0 \\
4823999931341641600 & 77.25054007 & -36.4610608 & 58739.3 & $ 15.96 \pm 0.05 $& 0.15 & 0103.C-0759 & HARPS & 0 \\
4823999931341640832 & 77.26015483 & -36.4644750 & 58738.3 & $ 14.25 \pm 0.32 $& 1.53 & 0103.C-0759 & HARPS & 1 \\
4823999931341640832 & 77.26015483 & -36.4644750 & 58719.4 & $ 13.34 \pm 0.36 $& 0.87 & 0103.C-0759 & HARPS & 1 \\
4823999931341640832 & 77.26015483 & -36.4644750 & 58740.3 & $ 17.64 \pm 0.28 $& 1.81 & 0103.C-0759 & HARPS & 1 \\
4823999931341640832 & 77.26015483 & -36.4644750 & 58701.4 & $ 16.03 \pm 0.56 $& 1.54 & 0103.C-0759 & HARPS & 1 \\
4823999931341640832 & 77.26015483 & -36.4644750 & 58720.4 & $ 15.11 \pm 0.31 $& 0.70 & 0103.C-0759 & HARPS & 1 \\
4823999931341640832 & 77.26015483 & -36.4644750 & 58750.3 & $ 16.21 \pm 0.82 $& 1.17 & 0103.C-0759 & HARPS & 1 \\
4823999931341640832 & 77.26015483 & -36.4644750 & 58689.4 & $ 16.51 \pm 0.36 $& 0.17 & 0103.C-0759 & HARPS & 1 \\

\hline
\hline
\end{tabular}%
}
\tablefoot{ We provide for each star the \textit{Gaia}-DR3 source identifier and coordinates, modified Julian date (MJD), radial velocity (and its uncertainty), radial velocity scatter computed from three different templates (as explained in Sect.~\ref{sect2_RV}), programme identifier of the spectrum, instrument and multiplicity flag (`0' = single star, `1'= binary or multiple system).}
\end{table*}

\begin{table*}[!h]
\centering
\caption{Properties of the 103 stars in our initial sample of the Octans young stellar association. This table will be available in its entirety at the CDS.
\label{tabA2}}
\resizebox{18cm}{!}{%
\begin{tabular}{lccccccccccc}
\hline\hline
Source identifier&$\alpha$&$\delta$&$\mu_{\alpha}\cos\delta$&$\mu_{\delta}$&$\varpi$&$RV_{GaiaDR3}$&$RV_{ThisPaper}$ &Flag&Sample&CS1&CS2\\
&(deg)&(deg)&(mas/yr)&(mas/yr)&(mas)&(km/s)&(km/s)\\
\hline\hline

4901711454686667008 & 6.27499723 & -62.6185534 & $ 32.415 \pm 0.033 $& $ 18.636 \pm 0.030 $& $ 11.936 \pm 0.028 $& $ 51.08 \pm 5.45 $& & 0 & 0 & 0 & 0 \\
4981632377929981056 & 15.31979557 & -45.9435677 & $ 34.872 \pm 0.012 $& $ 23.125 \pm 0.015 $& $ 13.387 \pm 0.018 $& $ 11.95 \pm 3.20 $& & 0 & 0 & 0 & 0 \\
4715353682706569728 & 20.15386593 & -62.6941181 & $ 33.460 \pm 0.036 $& $ 41.929 \pm 0.032 $& $ 14.516 \pm 0.028 $& & & 0 & 0 & 0 & 0 \\
4738546128147329536 & 40.32910404 & -57.4215814 & $ 17.962 \pm 0.055 $& $ 30.633 \pm 0.077 $& $ 11.513 \pm 0.060 $& $ 4.57 \pm 3.72 $& & 1 & 0 & 0 & 0 \\
4738546128148599296 & 40.32983626 & -57.4217087 & $ 14.865 \pm 0.023 $& $ 29.769 \pm 0.030 $& $ 11.404 \pm 0.024 $& $ -27.38 \pm 2.48 $& & 1 & 0 & 0 & 0 \\
4853696262938625920 & 51.90608334 & -37.9447831 & $ 6.978 \pm 0.040 $& $ 30.685 \pm 0.049 $& $ 13.295 \pm 0.042 $& & & 0 & 0 & 0 & 0 \\
4736642113310513792 & 53.66223637 & -49.9639832 & $ 9.362 \pm 0.078 $& $ 28.969 \pm 0.089 $& $ 10.798 \pm 0.068 $& & & 0 & 0 & 0 & 0 \\
4733231157659118464 & 55.05099827 & -52.1192680 & $ 0.895 \pm 0.018 $& $ 40.129 \pm 0.020 $& $ 12.965 \pm 0.016 $& $ 15.39 \pm 3.83 $& & 0 & 0 & 0 & 0 \\
4836484576636946560 & 57.62286029 & -43.6182879 & $ 6.059 \pm 0.015 $& $ 24.813 \pm 0.018 $& $ 9.393 \pm 0.014 $& $ 14.04 \pm 0.44 $& $ 14.75 \pm 0.26 $& 0 & 1 & 0 & 0 \\
4830136546254312576 & 61.86641950 & -50.7315293 & $ 0.132 \pm 0.059 $& $ 28.548 \pm 0.065 $& $ 9.952 \pm 0.051 $& & & 0 & 0 & 0 & 0 \\

\hline
\hline
\end{tabular}%
}
\tablefoot{We provide for each star the \textit{Gaia}-DR3 source identifier, position, proper motion, parallax, radial velocity from \textit{Gaia}-DR3, radial velocity derived in this study and binary flag (`0' = single star, `1' = binary or multiple star). The last three columns indicate whether the star is retained in our sample, CS1 and CS2 after applying the selection given in Sect.~\ref{sect2_sample_selection} (`0' = discarded, `1' = selected).}
\end{table*}

\begin{table*}[!h]
\centering
\caption{Youth indicators for the 103 stars in our initial sample of the Octans young stellar association. This table will be available in its entirety at the CDS.
\label{tabA3}}
\resizebox{18cm}{!}{%
\begin{tabular}{lcccccclc}
\hline\hline
Source Identifier& $G$ & $J$ & $NUV$ & HR1 & MIR excess & Li EW &  Youth indicators & Final member\\
&(mag)&(mag)&(mag)&& ($N\sigma$)  & ($\text{m}\AA$) & &\\
\hline\hline

4901711454686667008 & $ 15.289 $& $ 12.160 $& & & & &  & 0 \\
4981632377929981056 & $ 11.338 $& $ 9.447 $& $ 18.350 $& $ -0.025 $& $ 2.147 $& & UV;X;IR & 0 \\
4715353682706569728 & $ 15.497 $& $ 12.232 $& & & & &  & 0 \\
4738546128147329536 & $ 14.245 $& & $ 20.164 $& & & &  & 0 \\
4738546128148599296 & $ 12.788 $& $ 9.851 $& $ 20.164 $& & & & UV & 0 \\
4853696262938625920 & $ 16.292 $& $ 12.871 $& $ 22.956 $& & & & UV & 0 \\
4736642113310513792 & $ 17.465 $& $ 13.840 $& & & & &  & 0 \\
4733231157659118464 & $ 14.013 $& $ 11.114 $& & & & &  & 0 \\
4836484576636946560 & $ 10.350 $& $ 9.209 $& $ 15.777 $& $ -0.317 $& $ 2.832 $& $ 238 $& UV;IR;Li & 1 \\
4830136546254312576 & $ 15.986 $& $ 12.728 $& & & & &  & 0 \\

\hline
\hline
\end{tabular}%
}
\tablefoot{We provide for each star the \textit{Gaia}-DR3 source identifier, \textit{Gaia}-DR3 magnitude in the G-band, 2MASS magnitude in the J-band, NUV magnitude, HR1, significance of the MIR excess, lithium EW,  list of youth indicators and final membership in our sample. (`0' = discarded, `1' = selected). }
\end{table*}

\begin{table*}
\centering
\caption{Stellar positions and velocities for Octans members at present time ($t=0$) and birth time of the association ($t=-34$~Myr). This table will be available in its entirety at the CDS.
\label{tabA4}}
\resizebox{18cm}{!}{%
\begin{tabular}{lcccccccccccc}
\hline\hline
Source identifier&$X_{0}$& $Y_{0}$ & $Z_{0}$ & $U_{0}$ & $V_{0}$ & $W_{0}$ & $X$ & $Y$ & $Z$ & $U$ & $V$ & $W$\\
&(pc)&(pc)&(pc)&(km/s)&(km/s)&(km/s)&(pc)&(pc)&(pc)&(km/s)&(km/s)&(km/s)\\
\hline\hline

2888316685668937984 & $ -49.9 _{ -0.1 }^{+ 0.1 }$& $ -87.4 _{ -0.1 }^{+ 0.1 }$& $ -56.5 _{ -0.1 }^{+ 0.1 }$& $ -4.2 _{ -0.3 }^{+ 0.3 }$& $ 9.5 _{ -0.5 }^{+ 0.5 }$& $ -3.1 _{ -0.3 }^{+ 0.3 }$& $ 28.8 _{ -12.6 }^{+ 11.9 }$& $ -366.6 _{ -17.1 }^{+ 16.0 }$& $ 72.1 _{ -2.4 }^{+ 2.2 }$& $ 1.3 _{ -0.5 }^{+ 0.5 }$& $ 7.6 _{ -0.6 }^{+ 0.6 }$& $ 0.1 _{ -0.3 }^{+ 0.2 }$\\
2889149393928673664 & $ -59.4 _{ -0.1 }^{+ 0.1 }$& $ -106.8 _{ -0.2 }^{+ 0.2 }$& $ -57.9 _{ -0.1 }^{+ 0.1 }$& $ -4.9 _{ -0.3 }^{+ 0.3 }$& $ 8.2 _{ -0.5 }^{+ 0.5 }$& $ -4.0 _{ -0.3 }^{+ 0.3 }$& $ 40.3 _{ -12.6 }^{+ 12.4 }$& $ -338.9 _{ -17.8 }^{+ 17.3 }$& $ 79.9 _{ -2.0 }^{+ 2.0 }$& $ 0.5 _{ -0.5 }^{+ 0.5 }$& $ 6.1 _{ -0.6 }^{+ 0.6 }$& $ 0.7 _{ -0.2 }^{+ 0.2 }$\\
2892371925071938944 & $ -77.1 _{ -0.2 }^{+ 0.2 }$& $ -134.2 _{ -0.3 }^{+ 0.3 }$& $ -52.9 _{ -0.1 }^{+ 0.1 }$& $ -5.7 _{ -0.2 }^{+ 0.1 }$& $ 7.7 _{ -0.3 }^{+ 0.2 }$& $ -4.2 _{ -0.1 }^{+ 0.1 }$& $ 41.3 _{ -6.4 }^{+ 6.9 }$& $ -341.6 _{ -8.4 }^{+ 9.0 }$& $ 77.9 _{ -0.8 }^{+ 0.8 }$& $ 0.2 _{ -0.3 }^{+ 0.3 }$& $ 5.4 _{ -0.3 }^{+ 0.3 }$& $ 1.1 _{ -0.1 }^{+ 0.1 }$\\
2892719473825487104 & $ -67.8 _{ -0.1 }^{+ 0.1 }$& $ -119.2 _{ -0.2 }^{+ 0.3 }$& $ -54.8 _{ -0.1 }^{+ 0.1 }$& $ -5.3 _{ -0.1 }^{+ 0.1 }$& $ 8.1 _{ -0.1 }^{+ 0.1 }$& $ -4.3 _{ -0.1 }^{+ 0.1 }$& $ 42.3 _{ -3.3 }^{+ 3.4 }$& $ -343.2 _{ -4.3 }^{+ 4.7 }$& $ 80.0 _{ -0.5 }^{+ 0.5 }$& $ 0.3 _{ -0.1 }^{+ 0.1 }$& $ 5.9 _{ -0.2 }^{+ 0.1 }$& $ 1.1 _{ -0.1 }^{+ 0.1 }$\\
2893814896644875264 & $ -74.2 _{ -0.1 }^{+ 0.1 }$& $ -127.9 _{ -0.2 }^{+ 0.2 }$& $ -44.3 _{ -0.1 }^{+ 0.1 }$& $ -5.6 _{ -0.1 }^{+ 0.1 }$& $ 7.8 _{ -0.2 }^{+ 0.2 }$& $ -5.6 _{ -0.1 }^{+ 0.1 }$& $ 43.5 _{ -4.4 }^{+ 4.8 }$& $ -339.9 _{ -5.9 }^{+ 6.4 }$& $ 82.1 _{ -0.5 }^{+ 0.5 }$& $ 0.1 _{ -0.2 }^{+ 0.2 }$& $ 5.5 _{ -0.2 }^{+ 0.2 }$& $ 2.5 _{ -0.1 }^{+ 0.1 }$\\
2895580295703693056 & $ -91.2 _{ -0.2 }^{+ 0.2 }$& $ -141.4 _{ -0.3 }^{+ 0.3 }$& $ -48.7 _{ -0.1 }^{+ 0.1 }$& $ -5.2 _{ -0.5 }^{+ 0.4 }$& $ 9.0 _{ -0.7 }^{+ 0.7 }$& $ -4.4 _{ -0.3 }^{+ 0.2 }$& $ -3.7 _{ -19.4 }^{+ 21.6 }$& $ -396.4 _{ -24.0 }^{+ 26.3 }$& $ 76.2 _{ -1.8 }^{+ 1.9 }$& $ 1.6 _{ -0.9 }^{+ 0.8 }$& $ 7.2 _{ -0.9 }^{+ 0.9 }$& $ 1.3 _{ -0.2 }^{+ 0.2 }$\\
2898428000757810560 & $ -86.1 _{ -0.2 }^{+ 0.2 }$& $ -127.8 _{ -0.2 }^{+ 0.3 }$& $ -53.3 _{ -0.1 }^{+ 0.1 }$& $ -6.1 _{ -0.4 }^{+ 0.4 }$& $ 7.8 _{ -0.5 }^{+ 0.5 }$& $ -4.2 _{ -0.2 }^{+ 0.2 }$& $ 44.2 _{ -15.9 }^{+ 16.1 }$& $ -342.6 _{ -18.6 }^{+ 19.3 }$& $ 78.1 _{ -1.6 }^{+ 1.6 }$& $ -0.2 _{ -0.6 }^{+ 0.6 }$& $ 5.6 _{ -0.7 }^{+ 0.7 }$& $ 1.1 _{ -0.2 }^{+ 0.2 }$\\
2917020604947950976 & $ -90.0 _{ -0.3 }^{+ 0.3 }$& $ -99.0 _{ -0.3 }^{+ 0.3 }$& $ -49.9 _{ -0.2 }^{+ 0.1 }$& $ -6.7 _{ -0.2 }^{+ 0.3 }$& $ 7.5 _{ -0.3 }^{+ 0.3 }$& $ -5.5 _{ -0.1 }^{+ 0.1 }$& $ 71.5 _{ -10.7 }^{+ 9.8 }$& $ -310.7 _{ -9.9 }^{+ 9.3 }$& $ 85.3 _{ -1.0 }^{+ 1.0 }$& $ -1.5 _{ -0.4 }^{+ 0.4 }$& $ 5.4 _{ -0.3 }^{+ 0.4 }$& $ 2.2 _{ -0.1 }^{+ 0.1 }$\\
4772071921306849792 & $ -20.0 _{ -0.1 }^{+ 0.1 }$& $ -103.4 _{ -0.1 }^{+ 0.2 }$& $ -73.3 _{ -0.1 }^{+ 0.1 }$& $ -3.5 _{ -0.1 }^{+ 0.1 }$& $ 7.4 _{ -0.7 }^{+ 0.6 }$& $ -4.3 _{ -0.5 }^{+ 0.4 }$& $ 50.5 _{ -7.3 }^{+ 7.5 }$& $ -301.7 _{ -20.9 }^{+ 21.9 }$& $ 94.8 _{ -3.5 }^{+ 3.5 }$& $ 1.0 _{ -0.4 }^{+ 0.4 }$& $ 5.0 _{ -0.7 }^{+ 0.7 }$& $ 0.3 _{ -0.4 }^{+ 0.4 }$\\
4799008375640297344 & $ -32.7 _{ -0.1 }^{+ 0.1 }$& $ -103.9 _{ -0.2 }^{+ 0.2 }$& $ -71.0 _{ -0.1 }^{+ 0.1 }$& $ -3.8 _{ -0.1 }^{+ 0.1 }$& $ 8.4 _{ -0.4 }^{+ 0.5 }$& $ -3.0 _{ -0.3 }^{+ 0.3 }$& $ 39.1 _{ -7.6 }^{+ 7.2 }$& $ -339.6 _{ -15.8 }^{+ 14.9 }$& $ 83.4 _{ -2.5 }^{+ 2.2 }$& $ 1.3 _{ -0.3 }^{+ 0.4 }$& $ 6.1 _{ -0.5 }^{+ 0.5 }$& $ -0.5 _{ -0.3 }^{+ 0.2 }$\\

\hline
\hline
\end{tabular}%
}
\tablefoot{We provide for each star the \textit{Gaia}-DR3 source identifier, 3D positions and 3D velocities at present time ($X_{0},Y_{0},Z_{0},U_{0},V_{0},W_{0}$) and birth time of the association ($X,Y,Z,U,V,W$) with respect to the LSR. The uncertainties on the stellar positions and velocities were computed from a resampling procedure with 1\,000 samples generated from a Gaussian multivariate distribution of the observed parameters (positions, proper motions, parallaxes, and radial velocities) and the full covariance matrix of the data.}
\end{table*}

\end{appendix}
\end{document}